\documentclass[12pt,preprint]{aastex}

\newcommand{\msun}{M$_\odot$ }

\newcommand{\zaa}{A\&A~}

\newcommand{\zapj}{ApJ~}

\newcommand{\gap}{\mathrel{ \rlap{\raise.5ex\hbox{$>$}}
      {\lower.5ex\hbox{$\sim$}} } }
\newcommand{\lap}{\mathrel{ \rlap{\raise.5ex\hbox{$<$}}
		 {\lower.5ex\hbox{$\sim$}} } }

\newcommand{\al}{$^{26}$Al }

\begin{document}

\title{The Imprint of Nova Nucleosynthesis in Presolar Grains}

 \author{Jordi Jos\'e}
 \affil{Departament de F\'{\i}sica i Enginyeria Nuclear, Universitat Polit\`ecnica de 
  Catalunya, Av. V\'{\i}ctor Balaguer, s/n, E-08800 Vilanova i la Geltr\'u (Barcelona), Spain;}
 \affil{Institut d'Estudis Espacials de Catalunya (IEEC-UPC),
 Ed. Nexus-201, C/ Gran Capit\`a 2-4, E-08034 Barcelona, Spain; jordi.jose@upc.es}
 \author{Margarita Hernanz}
 \affil{Institut de Ci\`encies de l'Espai (CSIC), and Institut d'Estudis
 Espacials de Catalunya (IEEC-CSIC),
 Ed. Nexus-201, C/ Gran Capit\`a 2-4, E-08034 Barcelona, Spain; hernanz@ieec.fcr.es}
 \author{Sachiko Amari}
 \affil{Laboratory for Space Sciences and the Physics Department,
  Washington University, Campus Box 1105, One Brookings Drive, St. Louis, MO 63130-4899; sa@wuphys.wustl.edu}
 \author{Katharina Lodders}
 \affil{Planetary Chemistry Laboratory, Department of Earth \& Planetary, and McDonnell Center for the Space 
Sciences, Washington University, Campus Box 1169, St. Louis, MO 63130-4899; lodders@levee.wustl.edu}
 \and
 \author{Ernst Zinner}
 \affil{Laboratory for Space Sciences and the Physics Department,
  Washington University, Campus Box 1105, One Brookings Drive, St. Louis, MO 63130-4899; ekz@wuphys.wustl.edu}

\received{}
\accepted{}

\slugcomment{\underline{Submitted to}: \zapj ~~~~\underline{Version}:
\today}

\begin{abstract}

Infrared and ultraviolet observations of nova light curves have confirmed grain formation in their expanding 
shells that are ejected into the interstellar medium by a thermonuclear runaway. 
In this paper, we present isotopic ratios of intermediate-mass elements up to silicon for the ejecta 
of CO and ONe novae, based on 20 hydrodynamic models of nova explosions. These theoretical estimates 
will help to properly identify nova grains in primitive meteorites. In addition, equilibrium condensation 
calculations are used to predict the types of grains that can be expected in the nova ejecta, 
providing some hints on the puzzling formation of C-rich dust in O$>$C environments.  
These results show that SiC grains can condense in ONe novae, in concert with an inferred (ONe) nova origin 
 for several presolar SiC grains.

\end{abstract}

\keywords{novae, cataclysmic variables --- nucleosynthesis, abundances, isotopic anomalies, dust -- meteorites}

\section{Introduction}

 Classical novae are powered by thermonuclear runaways (TNRs) that occur on the white dwarf component (WD) 
of close binary systems (see Starrfield 1989, Kovetz \& Prialnik 1997, Jos\'e \& Hernanz 1998, Starrfield et al. 1998, 
and references therein).  During such violent stellar events, whose energy release is only exceeded by gamma-ray 
bursts and  supernova explosions, about $10^{-4}-10^{-5}$ $\rm M_\odot$ are ejected  into the interstellar medium. 
Because of the high peak temperatures attained during the outburst, $T_{peak} \sim (2-3) \times 10^8$ K, the ejecta 
are enriched in nuclear-processed material relative to the solar abundances, 
containing significant amounts of $^{13}$C, $^{15}$N, and $^{17}$O and 
traces of other isotopes, such as $^{7}$Li, $^{20}$Ne, $^{26}$Al, or $^{28}$Si (depending on the nova type,
CO or ONe, the mass of the underlying white dwarf, and other properties). Indeed, theoretical models of the
explosion reveal an isotopic pattern that does not correspond to equilibrium CNO burning (Starrfield et al. 1972).

In order to constrain the models, several studies have focused on a direct comparison of the atomic abundances, 
inferred from observations of the ejecta, with theoretical nucleosynthetic predictions (Jos\'e \& Hernanz 1998, 
Starrfield et al. 1998). Despite of problems associated with the modeling of the explosion (Starrfield 2002), 
such as the unknown mechanism responsible for the mixing of the accreted envelope and the outermost shells of 
the underlying white dwarf, or the difficulty of ejecting as much material as inferred from observations (see 
also Shore 2002), there is good agreement between theory and observations with regard to nucleosynthesis.
This agreement includes atomic abundance determinations (H, He, C, N, O, Ne, Na, Mg, Si...)
and a plausible endpoint for nova nucleosynthesis (around Ca).
For some well-observed novae, such as PW Vul 1984 or V1688 Cyg 1978, the agreement between observations and 
theoretical predictions (see Table 5, in Jos\'e \& Hernanz 1998, for details) is quite amazing. The reader is 
referred to Gehrz et al. (1998) for an extended list of abundance determinations in nova ejecta.

Moreover, since the nucleosynthesis path is very sensitive to details of the explosion 
(i.e., chemical composition, extent of convective mixing, thermal history of the envelope...), the agreement 
between the inferred abundances and the theoretical yields not only validates the {\it thermonuclear runaway model}, 
but also imposes limits on the nature of the mechanism responsible for the mixing. For instance, if one assumes 
that the mixing settles very late in the course of the explosion, pile-up of larger amounts of matter in 
the envelope would be favored since the injection of significant amounts of $^{12}$C, which triggers the onset 
of the TNR through proton-capture reactions, would be delayed. Hence, the explosion would take place in a somewhat 
more massive envelope, characterized by a higher ignition density (and pressure), giving rise to
 a more violent 
outburst with T$_{peak}$ exceeding in some cases $4 \times 10^8$ K (Starrfield; Jos\'e \& Hernanz;  
unpublished). Therefore, one would expect a significant enrichment in heavier species, 
beyond calcium, in the ejecta accompanying such violent outbursts. However, such an abundance pattern has never 
been seen in nature.

Nevertheless, a direct comparison with the elemental abundance pattern inferred from observations 
relies only on atomic abundances, and does not pose very strict limits on nova models. In contrast, a much more precise set of 
constraints may be obtained if information on specific isotopic abundances were available. One good example is 
silicon, with three stable isotopes (i.e., $^{28,29,30}$Si) in the region of interest for nova nucleosynthesis: 
whereas $^{28}$Si is strongly connected to the nature of the white dwarf core 
(either a CO or an ONe WD\footnote{The initial mass of the progenitor star determines the number of evolutionary
stages that it will undergo. Hence, stars within $2.3 \leq (M/M_\odot) \leq 8$ evolve through hydrogen and
helium burning, leaving a CO-rich white dwarf remnant. Stars in the mass interval 
$8 \leq (M/M_\odot) \leq 10-12$ additionally undergo carbon burning, leaving an ONe-rich remnant instead.}), 
 $^{29,30}$Si are good indicators of the peak temperatures achieved in the explosion and of the 
dominant nuclear paths followed in the course of the TNR, which have a clear imprint on the overall 
composition of the ejecta.

Such detailed information can be (partially) obtained through the laboratory analysis of presolar grains, which 
yields isotopic abundance ratios. Presolar grains, found in primitive meteorites, are characterized by huge isotopic 
anomalies that can only be explained in terms of nucleosynthetic processes that took place in their stellar sources.
In fact, detailed studies of these grains have opened up a new and promising field of astronomy (see Zinner 1998).
So far, silicon carbide (SiC), graphite (C), diamond (C), silicon nitride 
(Si$_{3}$N$_{4}$), and oxides (such as corundum and spinel) have been 
identified as presolar grains. Ion microprobe analyses of single presolar grains have revealed a variety of 
isotopic signatures that allow the identification of parent stellar sources, such as AGB stars and supernovae (Zinner 1998). 
Up to now, SiC grains have been most extensively studied and can be classified into different populations 
on the basis of their C, N, and Si isotopic ratios (see Figs. 1 and 2).

Infrared (Evans 1990, Gehrz et al. 1998, Gehrz 1999) and ultraviolet observations (Shore et al. 1994) 
of the evolution of 
nova light curves suggest that novae form grains in their expanding ejected shells. Both nova types, CO and ONe, 
behave in a similar way in the infrared immediately after the explosion, but as the envelope expands 
and becomes optically thin, differences in their infrared emission appear: whereas in a CO 
nova, this phase is typically followed by dust formation,  accompanied by a decline in visual 
brightness, and a simultaneous rise in infrared emission (see Rawlings \& Evans 2002; Gehrz 1999, 2002), 
ONe novae (erupting on more massive white dwarfs than CO novae) are not such prolific dust producers. 
The reason for this is that the latter have lower-mass, high-velocity ejecta, where the typical local 
densities may be too low to allow the condensation of appreciable amounts of dust. Observations of the 
condensation of dust containing different species, such as silicates, SiC, carbon and hydrocarbons have been 
reported for a number of novae (Gehrz et al. 1998).
The presence of SiC (or C-rich) dust in nova ejecta is established from spectroscopic measurements
(see Table 1, in Starrfield et al. 1998, and Table 2, in Gehrz et al. 1998). It is generally believed that 
C$>$O is needed for the formation of SiC and/or graphite grains. If oxygen is more abundant than carbon, essentially 
all C is locked up in the very stable CO molecule, and the excess O leads to formation of oxides 
and silicates as condensates. On the other hand, if carbon is more abundant that oxygen, essentially all O is tied up in CO 
and the excess C can form reduced condensates such as SiC or graphite. 
Since theoretical models of nova outbursts yield, on average,  O$>$C, one would expect 
only oxidized condensates using the carbon and oxygen abundances as a sole criterion. However, this is at odds with 
the observation of C-rich dust detected around some novae (Gehrz, Truran \& Williams 1993; Starrfield et al. 1997; Gehrz 1999).

While previously the identification of presolar nova grains from meteorites relied only on low $^{20}$Ne/$^{22}$Ne 
ratios (with $^{22}$Ne being attributed to $^{22}$Na decay; Amari et al. 1995; Nichols et al. 2004),  
recently five SiC and two graphite grains that exhibit other isotopic signatures characteristic of nova nucleosynthesis 
have been identified (see Hoppe et al. 1995, Amari et al. 2001, Amari 2002, 
for details). This discovery provides a very valuable set
of constraints for nova nucleosynthesis. 
Table 1 summarizes the mineralogy and isotopic composition
of these grains, reported in Amari et al. (2001) and Amari (2002). 
The SiC grains have very low $^{12}$C/$^{13}$C and $^{14}$N/$^{15}$N ratios, while the graphite grains have 
low $^{12}$C/$^{13}$C, but normal $^{14}$N/$^{15}$N ratios. However, the original $^{14}$N/$^{15}$N ratios of 
these two graphite grains could have been much lower, because there is evidence that indigenous N in presolar 
graphites has been isotopically equilibrated with terrestrial nitrogen. For example, most presolar graphite 
grains show a huge range in C isotopic ratios but essentially normal (terrestrial) N isotopic composition 
(Hoppe et al. 1995). Recent isotopic imaging of C and O inside of slices of graphite spherules showed gradients 
from highly anomalous ratios in the center to more normal ratios close to the surface, also indicating 
isotopic equilibration (Stadermann et al. 2004). $^{26}$Al/$^{27}$Al ratios have been determined 
only for two SiC grains (KJGM4C-100-3 and KJGM4C-311-6) and are very high ($> 10^{-2}$; see Fig. 3).
We note that the $^{20}$Ne/$^{22}$Ne ratio is only available for the graphite grain KFB1a-161 ($ < 0.01$;
 $^{22}$Na/C = $9 \times 10^{-6}$; see Nichols et al. 2004), being considerably lower than the ratios 
predicted by nova models (see Section 2). Usually, neon is incorporated in grains via implantation, 
since noble gases do not condense as stable compounds into grains (Amari 2002). However, the 
low $^{20}$Ne/$^{22}$Ne ratio measured in this grain suggests that Ne has not been implanted 
in the ejecta, but $^{22}$Ne most likely originated from in situ decay of
$^{22}$Na (with a mean lifetime $\tau$=3.75 yr).

Silicon isotopic ratios of the five SiC grains are characterized by $^{30}$Si excesses and close-to- or 
slightly lower-than-solar $^{29}$Si/$^{28}$Si ratios. Whereas CO nova models (Kovetz \& Prialnik 1997; 
Starrfield et al. 1997; Jos\'e \& Hernanz 1998, Hernanz et al. 1999, and unpublished data) predict close-to-solar 
$^{30}$Si/$^{28}$Si and close-to- or lower-than-solar $^{29}$Si/$^{28}$Si, huge enrichments of $^{30}$Si and 
close-to or lower-than-solar $^{29}$Si/$^{28}$Si ratios are obtained for ONe novae (Jos\'e \& Hernanz 1998; 
Hernanz et al. 1999, Jos\'e, Coc, \& Hernanz 1999, 2001; Starrfield et al. 1998).
We have also included unpublished data on Si isotopic ratios for grain KFB1a-161, both in 
Table 1 and in Figure 2. Unfortunately, trace element concentrations in KFB1a grains are low and hence,
measurements are characterized by large errors.

The isotopic signatures of these grains qualitatively agree with current predictions from hydrodynamic models 
of nova outbursts. In fact, a comparison between grain data and nova models suggests that these grains  
formed in ONe novae with a white dwarf mass of at least 1.25M$_{\odot}$ (Amari et al. 2001). However, two
main problems, related with the likely nova paternity of these grains, remain to be solved yet: first, the
challenging connection with ONe novae which, as stated before, are not as prolific dust producers than CO
novae,  and second, 
 in order to quantitatively match the grain data, one has to assume a mixing process between 
 material newly synthesized in the nova outburst and  more than ten times as much unprocessed, isotopically 
 close-to-solar, material before grain formation.

In this paper, we provide theoretical predictions for the expected isotopic composition of 
the nova ejecta, and explore which type of condensates may form. The structure of the paper is as follows: in Section 2, 
we summarize the main nucleosynthesis results from one-dimension, hydrodynamic computations of nova outbursts. We 
report the theoretically expected isotopic ratios in the ejected nova shells, which should also be representative
of the isotopic composition of grains condensed in the ejecta. Section 3 describes the results of chemical 
equilibrium condensation calculations for different types of nova ejecta. In Section 4, we explore 
the isotopic patterns of individual ejected shells and compare them with those resulting from mean mass-averaged 
envelopes. Furthermore, we analyze the effect of relevant nuclear physics uncertainties on the results presented here,
and compare them with results obtained by other groups. 
A summary of the main conclusions of this paper is given in Section 5.

\section{Theoretical isotopic ratios in nova ejecta: Mean mass-averaged values}

We have adopted two different approaches in our search for trends in the
isotopic composition
that may characterize the nova ejecta: first, an analysis based on mean
mass-averaged isotopic ratios for a number of species, resulting
from hydrodynamic calculations of classical nova
outbursts, and second, a detailed analysis of the chemical
abundance gradients found when individual ejected 
shells are taken into account. It is important to stress that mean 
mass-averaged ratios provide a global 
view of the nucleosynthetic imprints of the explosion. 
The envelopes ejected in our
numerical models of the nova outburst consist 
of a large number of shells of different masses (which decrease outwards
from the envelope's base). The innermost shells are
probably the most relevant ones, the reason being twofold:
first, these are the shells that undergo the largest changes in chemical
composition through nuclear processing (i.e., the shells that will exhibit
the strongest imprints of a nova outburst) and second, because of their
larger masses, material from these shells has a larger probability to condense and form dust 
and grains.
Both aspects are partially taken into account by the mass-averaging  
process\footnote{In fact, the analysis available in
Starrfield et al (1997), the only work that addresses a similar search of
nova nucleosynthesis trends and its connection with meteorites, is exclusively
based on mean mass-averaged isotopic ratios} 
that assigns different weights to individual shells.
In contrast, a quantitative analysis based on individual shells, 
although in principle more detailed, can be potentially misleading: it may
generate a biased view of the nucleosynthetic history since, a priori,
all possible ratios found throughout the envelope (see Tables 2 \& 3, and 
variation bars in Figs. 4 to 9)
seem, at a first glance, equally likely. It is therefore important
to point out that the largest deviations from the mean are often
obtained in individual shells located near the surface, in low mass shells 
with a lower probability of forming grains and with isotopic features 
that reflect to a much lower extent the imprint of a nova outburst.
In fact, the differences found 
in those surface layers are connected with details of the retreat of 
convection from the surface rather than caused by 
nuclear processes. It is therefore our aim to focus this analysis first
on mean mass-averaged ratios and then to address the question of how robust our conclusions are
when an analysis based on individual shells is performed (Section 4.1).

A full list of mean mass-averaged values, together with maximum and minimum 
isotopic ratios throughout the ejecta, for a sample of 20 
hydrodynamic nova models, is given in Tables 2 \& 3. 
Details of the numerical code, developed to follow the course of 
nova outbursts, from the onset of accretion  to the expansion and ejection stages, have been 
discussed by Jos\'e \& Hernanz (1998), 
and are summarized in the accompanying Appendix, together with a brief
description of the initial isotopic ratios adopted in the models presented here.

\subsection{Nitrogen and carbon isotopic ratios}

The final $^{14}$N/$^{15}$N ratios found in the ejected shells of nova
outbursts show a wide range of variation (see Fig. 4). Explosions involving ONe white dwarfs 
yield low ratios, ranging from $\sim$ 0.3 to 4 (solar ratio = 272). In contrast, CO nova models are 
characterized by higher ratios, typically between $\sim$ 3 and 100 (see Fig. 4), but as high as $\sim 1400$ for the 
extreme 0.6 \msun CO case. As we will stress throughout this Section, the nuclear activity in this low-mass CO model 
is so tiny that the final isotopic ratios are, for many species, close to the initial ratios of the envelope at the 
onset of the TNR.
These differences in the final N ratios between CO and ONe models reflect 
differences in the main nuclear paths followed in the course of the explosions. 

The synthesis of $^{15}$N depends critically on the amount of $^{14}$N available (both the initial one, present in 
the accreted material, as well as the amount synthesized through the CNO cycle, starting from $^{12}$C). Since both CO 
and ONe models begin with the same initial $^{14}$N
 (see Table 3),
differences in the ejecta reflect different thermal histories 
during the explosion (and in particular, differences in $T_{peak}$): the higher peak temperatures achieved in ONe models 
favor proton-capture reactions on $^{14}$N, leading to $^{14}$N(p,$\gamma$)$^{15}$O($\beta^+$)$^{15}$N, and are thereby
responsible for the higher $^{15}$N content in the ejecta.
This explains also why the $^{14}$N/$^{15}$N ratio decreases as the white dwarf mass increases, for both CO and ONe 
models, a direct consequence of the higher temperatures achieved for more massive white dwarfs. 
In summary, the $^{14}$N/$^{15}$N ratio provides a means for distinguishing between CO and ONe novae: large ratios, 
of the order of 100-1000, are only achieved in explosions involving low-mass CO novae, according to the models discussed.

In contrast to the N isotopic ratios, both CO and ONe models yield very 
low $^{12}$C/$^{13}$C ratios (see also Fig. 4), in 
the range $\sim$ 0.3-2 (solar ratio = 89). 
The dramatic 
reduction in the final $^{12}$C/$^{13}$C ratios as compared to the initial ones is due to the very efficient synthesis 
of $^{13}$C through $^{12}$C(p,$\gamma$)$^{13}$N($\beta^+$)$^{13}$C which, in turn, decreases the final amount of $^{12}$C.

The effect of the white dwarf mass on the $^{12}$C/$^{13}$C ratios follows also a certain pattern, but unlike the case 
of N isotopic ratios, it depends on the nova type: for low-mass CO white dwarfs, the amount of $^{13}$C synthesized from 
proton-capture reactions on $^{12}$C is strongly limited by the moderate range of temperatures achieved in the explosion. 
However, as the mass of the white dwarf (and hence, the temperature at the envelope's base) increases, more $^{13}$C 
is produced, leading to lower $^{12}$C/$^{13}$C ratios, up to a point where the temperatures achieved in the envelope 
are high enough to enable significant proton-captures on $^{13}$C, increasing again the $^{12}$C/$^{13}$C ratio. 
In ONe novae, the temperatures achieved during the explosions are always high enough for significant 
proton-capture reactions to proceed on $^{13}$C (which, in turn, increase the final amount of $^{14}$N), leading to 
$^{12}$C/$^{13}$C ratios that monotonically increase with the white dwarf mass.

We stress that C ratios are highly diagnostic for identifying potential nova grain candidates since,
as can be seen in Fig. 4, independently of the nova type and of the adopted white dwarf mass, all models are
characterized by an extremely narrow range of low $^{12}$C/$^{13}$C values (in contrast with the wide dispersion
obtained for the N ratios), definitely a characteristic signature of a classical nova explosion.

\subsection{Oxygen and neon isotopic ratios}

Oxygen isotopic ratios depend on the nova type (i.e., CO or ONe) and on the white dwarf mass. As can be seen in Fig. 5, 
CO models are in general characterized by moderate to large 
$^{16}$O/$^{18}$O ratios, ranging from 20 to 39,000 (solar ratio = 498), 
and moderate $^{16}$O/$^{17}$O ratios, from 
 8 to 230 (solar ratio = 2622). In contrast, lower ratios are, in general, found for ONe models: 
whereas $^{16}$O/$^{18}$O ranges from 10 to 400, $^{16}$O/$^{17}$O ranges from 1 to 10. It is important to stress 
that a recent revision of the $^{18}$F(p,$\alpha$) reaction rate 
(see Hernanz et al. 1999, and Coc et al. 2000) yields, in general, higher $^{16}$O/$^{18}$O ratios than those obtained 
with previous estimates (compare, for instance, the O ratios obtained
in models ONe3 and ONe4). 
At the time the calculations presented in this paper were made, the 
$^{18}$F(p,$\alpha$) rate was affected by a large uncertainty (a factor of $\sim 300$). Two recent experiments,
at Oak Ridge and Louvain-la-Neuve, have reduced this uncertainty by a factor of $\sim 5$, but clearly, a significant 
uncertainty still remains. However, it is important to stress that, while the uncertainty in the rate has been reduced
there is no clear indication of a significant deviation from the nominal rate that was used in our calculations, 
thus our conclusions concerning the $^{16}$O/$^{18}$O ratios remain essentially unaffected. 

The decrease from the huge initial ratios (see Table 4) down to the values predicted for the ejecta is a measure of the nuclear 
processes that transform $^{16}$O to $^{17,18}$O, beginning with proton captures onto $^{16}$O, which require high 
enough temperatures to overcome the large Coulomb potential barrier. Indeed, at the typical temperatures attained in nova outbursts, 
the abundance of $^{16}$O always decreases, since $^{16}$O(p,$\gamma$) dominates over $^{15}$N(p,$\gamma$)$^{16}$O, 
$^{19}$F(p,$\alpha$)$^{16}$O, and 
$^{17}$F($\gamma$,p)$^{16}$O. $^{17}$O is 
synthesized by $^{16}$O(p,$\gamma$)$^{17}$F($\beta^+$)$^{17}$O, and can be destroyed either by 
$^{17}$O(p,$\gamma$)$^{18}$F (which decays into $^{18}$O) or by $^{17}$O(p,$\alpha$)$^{14}$N. The dominant 
destruction reaction for $^{18}$O is $^{18}$O(p,$\alpha$). Since ONe models, which
reach higher peak temperatures than CO models, 
synthesize larger amounts of both $^{17,18}$O, they are characterized by lower $^{16}$O/$^{17}$O ratios and 
similar or lower $^{16}$O/$^{18}$O ratios than models of CO novae.

A similar trend is found when looking at the role of the white dwarf mass (Fig. 5): in CO models, as the mass of 
the white dwarf increases (accompanied by increasing temperatures in the envelope), the $^{16}$O/$^{17}$O and 
$^{16}$O/$^{18}$O ratios both decrease. No trend is clearly seen for ONe novae: the synthesis of 
$^{17}$O has a maximum at a temperature around $2 \times 10^8$ K (i.e., 1.25 \msun ONe models), however, because of 
the decrease in $^{16}$O, the $^{16}$O/$^{17}$O ratio decreases in general as the mass of the white dwarf 
increases. Most ONe models show similar $^{16}$O/$^{18}$O ratios, around
 $\sim$ 200 - 400 with the new $^{18}$F(p,$\alpha$) rates (see Hernanz et al.
 1999), with no clear dependence on the white dwarf mass.

Another interesting source of information are the neon isotopic ratios. They are useful for distinguishing 
between CO and ONe novae: the higher initial $^{20}$Ne content in ONe novae 
is the main reason for the much higher $^{20}$Ne/$^{22}$Ne ratios found in those models (Fig. 6), ranging typically from 
$\sim 100$ to $250$ (solar ratio = 14). 
In contrast, CO models yield $^{20}$Ne/$^{22}$Ne ratios ranging only from $\sim 0.1$ to $0.7$.

Differences between CO and ONe novae are not so extreme with regard to the $^{20}$Ne/$^{21}$Ne ratio (see also Fig. 6), 
and in fact, ratios for the two nova types overlap at values of 
$\sim$ 2500 - 10,000. 
The increase in the $^{20}$Ne/$^{21}$Ne ratio with respect to the initial
value (see Table 4 and Appendix) reflects the fact that 
$^{20}$Ne is scarcely modified in most nova models, since its destruction by proton-capture reactions requires rather 
high temperatures. On the other hand, $^{21}$Ne, a fragile isotope, is almost completely destroyed: first, by proton-capture reactions 
and, as the temperature rises, the synthesis path through $^{20}$Ne(p,$\gamma$)$^{21}$Na($\beta^+$)$^{21}$Ne 
is halted as soon as proton-captures on $^{21}$Na become faster than its $\beta^+$-decay. This accounts for the large 
$^{20}$Ne/$^{21}$Ne ratios found in both nova types.

\subsection{Aluminum and magnesium isotopic ratios}

Similar $^{26}$Al/$^{27}$Al ratios (typically, $\sim 0.01 - 0.6$. See Fig. 7) 
are obtained for both CO and ONe nova models.
Although $^{26}$Al is efficiently synthesized only in ONe novae, the larger initial amount of $^{27}$Al in such novae,  
more than two orders of magnitude higher than in CO novae (and not strongly modified during the explosion), results in 
similar ratios in the two nova types. Therefore, the $^{26}$Al/$^{27}$Al ratio is not a diagnostic for distinguishing 
between CO and ONe novae.
Nucleosynthesis of \al is complicated by the presence of a short-lived $^{26}$Al$^m$ ($\tau$=9.15~s) spin isomer. 
The only way to synthesize the long-lived $^{26}$Al$^g$ isotope ($\tau=1.04 \times 10^6$~yr) in nova explosions is 
through proton-capture reactions on ${}^{25}$Mg, which can yield both the $^{26}$Al ground and isomeric states. 
Hence, the $^{25}$Mg abundance is critical for the synthesis of $^{26}$Al. Eventually, other isotopes, such as 
$^{24}$Mg, $^{23}$Na and, to some extent, $^{20}$Ne, can also contribute to the final \al yield (see Jos\'e, 
Coc \& Hernanz 1999). The synthesis of $^{27}$Al is also complicated: whereas it is mainly destroyed by 
$^{27}$Al(p,$\gamma$), several mechanisms compete in its synthesis: one is $^{26}$Mg(p,$\gamma$), with $^{26}$Mg 
coming from its initial abundance as well as from $^{26}$Al$^m$-decay (synthesized by $^{25}$Mg(p,$\gamma$) or
through two proton-captures on $^{24}$Mg, leading to the $\beta^+$-unstable $^{26}$Si); another possibility is 
$^{27}$Si($\beta^+$)$^{27}$Al, with $^{27}$Si coming from both $^{26}$Al$^{g,m}$(p,$\gamma$). There is some tendency to obtain higher $^{26}$Al/$^{27}$Al ratios for more massive white dwarfs. 

Both CO and ONe nova 
models yield, in general, low $^{24}$Mg/$^{25}$Mg ($\sim 0.02-0.3$) and $^{26}$Mg/$^{25}$Mg ($\sim 0.07-0.2$) 
 ratios (see Fig. 8), except for the extreme 0.6 \msun CO case, with $^{24}$Mg/$^{25}$Mg = 4.3 and 
$^{26}$Mg/$^{25}$Mg = 0.7. 
The CO nova models show a rather complicated pattern, because Mg synthesis is very sensitive to the maximum temperature 
(and hence, to the adopted white dwarf mass) attained in the explosion. Since proton captures on $^{26}$Mg
  require high enough temperatures to overcome its large Coulomb barrier, the final $^{26}$Mg abundances are, in general, 
very close to the initial ones (with only a small decrease for the 1.15 \msun CO models). Again, the 0.6 \msun CO model 
shows no imprint of nuclear activity involving $^{24}$Mg, but already the 0.8 \msun CO model begins to show a
  decrease in the final $^{24}$Mg yield, since at the moderate temperatures reached in this model 
$^{24}$Mg(p,$\gamma$)$^{25}$Al dominates $^{23}$Na(p,$\gamma$)$^{24}$Mg.  This in turn explains the increase
 in $^{25}$Mg powered by $^{25}$Al($\beta^+$)$^{25}$Mg. However, when the temperature reaches $\sim 2 \times 10^8$ K 
(as for the 1.15 \msun CO models), the rates for $^{24}$Mg(p,$\gamma$)$^{25}$Al and $^{23}$Na(p,$\gamma$)$^{24}$Mg
 become comparable, and hence the decrease in the final $^{24}$Mg yield is halted. At the same time, $^{25}$Mg(p,$\gamma$)
  dominates over $^{24}$Mg(p,$\gamma$)$^{25}$Al($\beta^+$)$^{25}$Mg, which accounts for some decrease in the $^{25}$Mg yield.

  In contrast, for the ONe models the final Mg yields do not depend much on the adopted white dwarf mass: in all
   models, both the final $^{24}$Mg and $^{26}$Mg abundances 
   are significantly lower than the initial values (by two and one orders 
   of magnitude, respectively), whereas $^{25}$Mg decreases by only a factor of $\sim$ 2.
  Most of the destruction of Mg isotopes takes place at temperatures around $2 \times 10^8$ K (see Jos\'e et al. 1999). 
 The differences with respect to the results for CO models are essentially due to significant differences in the initial 
chemical composition (the ONe models are, for instance, much richer in $^{23}$Na and $^{25,26}$Mg),
  which affects not only the dominant nuclear path, but also the characteristic timescales of the explosion and hence the 
exposure time to potential proton-capture reactions.

\subsection{Silicon isotopic ratios}

CO novae show, in general, a very limited nuclear activity beyond the CNO mass region, because of the moderate peak 
temperatures attained during the explosion and also because of the lack of significant amounts of 'seed' nuclei above 
this mass range. Therefore, hydrodynamic models of CO novae yield close-to-solar Si isotopic ratios in the ejecta. 
Only the most massive CO models (i.e., 1.15 \msun) show marginal activity in the Si region, powered by a moderate leakage 
from the MgAl region through $^{26}$Al$^g$(p,$\gamma$)$^{27}$Si, followed either by 
$^{27}$Si(p,$\gamma$)$^{28}$P($\beta^+$)$^{28}$Si, or by $^{27}$Si($\beta^+$)$^{27}$Al(p,$\gamma$)$^{28}$Si, which 
compete favorably with $^{28}$Si(p,$\gamma$), and hence tend to increase the amount of $^{28}$Si. In contrast, 
the mass fraction of $^{29}$Si decreases, since destruction through $^{29}$Si(p,$\gamma$) dominates synthesis by 
$^{28}$Si(p,$\gamma$)$^{29}$P($\beta^+$)$^{29}$Si. The $\beta^+$-decay of the residual $^{30}$P nuclei is responsible 
for some marginal overproduction of $^{30}$Si.

Silicon isotopic ratios are usually expressed as 
$\delta ^{29,30}$Si/$^{28}$Si = [($^{29,30}$Si/$^{28}$Si)/($^{29,30}$Si/$^{28}$Si)$_\odot$ - 1] $\times$ 1000, 
which represent deviations from solar abundances in permil. 
As shown in Fig. 9, 
all CO models are characterized by close-to or lower-than-solar $\delta ^{29}$Si/$^{28}$Si, and by close-to-solar 
$\delta ^{30}$Si/$^{28}$Si. 
A quite different pattern is found for the ONe models, partially because of the higher peak temperatures achieved 
during the explosion, but also because of the higher initial $^{27}$Al abundance. The abundance of $^{28}$Si increases 
from 1 to 1.25 \msun ONe models and then decreases a bit for 1.35 \msun models. This results from the fact that around 
T=$10^8$ K, $^{27}$Al(p,$\gamma$)$^{28}$Si dominates $^{28}$Si(p,$\gamma$). When the temperature rises, 
$^{26}$Al$^{m,g}$(p,$\gamma$)$^{27}$Si(p,$\gamma$)$^{28}$P($\beta^+$)$^{28}$Si also contributes to $^{28}$Si synthesis, 
but as the temperature reaches $\sim 3 \times 10^8$ K, destruction through proton-capture reactions dominate all 
reactions leading to $^{28}$Si synthesis. Hence, there is a maximum in the $^{28}$Si production around 1.25 \msun ONe models 
(which attain $T_{peak} < 3 \times 10^8$ K).

In contrast, both $^{29,30}$Si increase monotonically with the white dwarf mass. They are powered by 
$^{29,30}$P($\beta^+$)$^{29,30}$Si, which dominate destruction through proton-capture reactions. Fig. 9 shows an increase 
in $\delta ^{30}$Si/$^{28}$Si with increasing white dwarf mass: whereas 1.0 \msun ONe models show a noticeable destruction 
of $^{30}$Si, 1.15 \msun ONe models yield close-to-solar $\delta ^{30}$Si/$^{28}$Si values. Excesses appear for 
$M_{wd} \geq 1.25$ \msun, as a result of the higher temperatures attained in the envelope. On the other hand, 
$\delta ^{29}$Si/$^{28}$Si ratios are below solar and only approach close-to-solar values when the white dwarf mass
 reaches 1.35 M$_\odot$.

\subsection{Isotopic ratios of elements beyond silicon}

For classical novae, nuclear activity above silicon is limited to events involving a very massive ONe white dwarf, 
close to the Chandrasekhar limit, since nucleosynthesis in the Si-Ca mass region requires temperatures above 
$3 \times 10^8$ K to overcome the large Coulomb barriers of those elements (Politano et al. 1995; Jos\'e \&
Hernanz 1998; Starrfield et al. 1998).  Indeed, observations 
of some novae reveal the presence of nuclei in this mass range in their spectra, including sulfur 
(Nova Aql 1982, Snijders et al. 1987, Andre\"a et al. 1994), chlorine (Nova GQ Mus 1983, Morisset \& Pequignot 1996), 
argon and calcium (Nova GQ Mus 1983, Morisset \& Pequignot 1996; Nova V2214 Oph 1988, Nova V977 Sco 1989 and 
Nova V443 Sct 1989, Andre\"a et al. 1994). Models of explosions on 1.35 \msun ONe white dwarfs yield large overproduction 
factors (i.e., $f = X_i/X_{i,\odot}$) for a number of isotopes, including $^{31}$P ($f \sim 1100$), $^{32}$S ($f \sim 110$), $^{33}$S ($f \sim 150$), 
and $^{35}$Cl ($f \sim 80$) (Jos\'e et al. 2001).  
However, the chances to measure such 
excesses in presolar grains are scarce. Although the predicted $^{33}$S excess may provide a remarkable signature of 
a classical nova event, no sulfur isotopic measurements have been made so far on presolar SiC grains. Nevertheless, 
equilibrium condensation calculations predict that sulfides might be incorporated into SiC grains (Lodders \& Fegley 1995).
 Analyses of graphite grains (unpublished data) yielded solar S isotopic ratios, and it is likely that the 
measured S is dominated by contamination or that any indigenous S has been isotopically equilibrated during the 
chemical separation procedure. Since $^{31}$P is the only stable phosphorus isotope, no P isotopic ratios can be obtained. 
One chance might be to measure the ratio of two isotopes from different elements (such as P and S), but this would require
 information on the condensation behavior of these elements, which usually cannot be obtained. Two stable isotopes are 
available for chlorine, but there is not much of a chance to measure excesses in $^{35}$Cl because Cl is not expected 
to condense into SiC. Furthermore, the standard separation procedure for SiC uses HCl, introducing a strong Cl contamination.

\section{Formation of grains}

We explored grain formation by calculating thermodynamic equilibrium condensation sequences of the 
 ejected layers for three representative nova models, involving 1.15 \msun CO, 1.15 \msun ONe and
1.35 \msun ONe WD. In particular, we adopted the chemical composition of the first, innermost ejected layer in each
case, since  the largest changes in chemical composition from solar are expected to be found precisely in the innermost shells 
of the envelope. We note that similar results are found if nearby shells (i.e., 5 or 10 shells above the innermost
ejected one) are used instead, but towards the outermost shells chemical variations are 
less extreme and different condensates are likely to appear. The isotopic chemical pattern of those envelope shells,
calculated self-consistently by means of the hydrodynamic code (for nuclei ranging between H and Ca), was augmented by 
assuming solar element/Ca abundance ratios for elements heavier than Ca. Calculations were performed with the {\it CONDOR} 
code and the computational procedure is similar to that described 
in Lodders \& Fegley (1995, 1997) and Lodders (2003). The calculations use the temperature and 
pressure profile computed for the whole envelope with the hydrodynamic code. At temperatures where condensation occurs, total 
pressures are in the range of $10^{-6}$ to $10^{-7}$ bar. 
It should be stressed that the results of these computations only apply to the innermost ejected shell of the expanding 
nova ejecta and the underlying assumption is that  no mixing occurs between this and overlying shells. Therefore, the 
calculations only describe the condensates for an extreme endmember composition of the overall ejecta from a given nova model, 
and additional condensates of different mineralogy that may be produced in the outer shells are not considered here. The 
investigation of the condensates that can form in the different ejecta layers and in overall homogenized ejecta will be 
described elsewhere (Lodders et al. 2004).

\subsection {Condensates for 1.15 \msun CO nova ejecta}

The atomic C/O ratio of the innermost shell of this CO nova model is $\sim 0.8$, and we expected oxides and silicates as 
condensates, similar to those that condense from a solar-composition gas. This expectation was met and the condensation sequence 
is shown in Fig. 10 (upper panel).  The first 
condensate is corundum (Al$_2$O$_3$) at 1743 K, followed by hibonite (CaAl$_{12}$O$_{19}$) at 1567 K. 
Gehlenite (Ca$_2$Al$_2$SiO$_7$), the Al-rich endmember of the gehlenite-akermanite solid solution called melilite, appears in 
addition at 1469 K. This phase consumes Ca, which is much less abundant than Al. This limits the stability of hibonite, 
and corundum is stable again after melilite appears. With further decrease in temperature, melilite transforms into anorthite 
(CaAl$_2$Si$_2$O$_8$), and corundum into spinel (MgAl$_2$O$_4$). Cordierite (Mg$_2$Al$_4$Si$_5$O$_{18}$) appears together with 
anorthite and spinel below $\sim 1280$ K. In fact, cordierite is not a stable condensate in a solar composition gas, but 
appears here because of the larger relative abundances of Mg, Si, and Al. Cordierite eventually consumes more Al so that 
the spinel stability 
is terminated. Substantial removal of Si and Mg from the gas starts when forsterite (Mg$_2$SiO$_4$) and then enstatite 
(MgSiO$_3$) condense near 1300 K and, in addition to enstatite, SiO$_2$ appears as a separate phase.
The first Ti-bearing condensate is a calcium titanate (Ca$_4$Ti$_3$O$_{10}$) at 1500 K, which converts to perovskite 
(CaTiO$_3$) at lower temperatures. Anorthite is the major sink for Ca and perovskite transforms into Ti$_4$O$_7$ shortly after 
anorthite becomes stable. Metallic iron condenses at 1166 K and phosphorus condenses as Fe$_3$P at 1053 K. Iron sulfide (FeS) only 
forms at low temperatures of $\sim 720$ K.
In this condensation sequence, corundum, spinel and enstatite are the only minerals which are also found among the major
presolar grain types. 
Reduced condensates such as SiC or graphite do not appear, which suggests that models similar to this, involving a 
1.15 \msun CO WD, will not contribute to the presolar SiC or graphite grains.

\subsection {Condensates for 1.15 \msun ONe nova ejecta}

The C/O ratio of the first, innermost ejected shell  is now $\sim 0.7$, 
below unity and only slightly lower than that of the CO nova model of the same mass. If the C/O ratio were the sole criterion, 
oxidized condensates would be expected as well. However, by comparison to solar, the abundances of Al, Ca, Mg and Si are 
fairly high in ONe ejecta, which means that the C and O chemistries will be affected by the abundances of the rock-forming 
elements. Fig. 10 (middle panel) shows the calculated condensate stabilities as a function of temperature. The first condensate is 
corundum at 1644 K, followed by AlN and TiN. The AlN only coexists with corundum for a short temperature interval, and 
oxidized Al-bearing compounds (i.e., hibonite, melilite, anorthite) coexist with corundum instead at lower temperatures. 
Near 1120 K, corundum turns into andalusite (Al$_2$SiO$_5$) and cordierite when more of the abundant silicon 
is removed from the gas.
The TiN stability range is interrupted for a brief temperature step from $\sim 1230$ to 1250 K, when TiC is more stable, but 
the TiC stability is limited by the appearance of SiC at 1235 K. The first Si-bearing condensate is FeSi, and SiC is the 
next stable one. Sinoite (Si$_2$N$_2$O) enters the suite of condensates at 1110 K, followed by enstatite around 1100 K 
and SiO$_2$ at $\sim 1090$ K. With the appearance of sinoite and enstatite the SiC stability ends.
The occurrence of SiC, corundum, and enstatite in this ejected shell suggests that such intermediate-mass 
ONe novae could contribute to the known 
presolar SiC, corundum, and enstatite grain populations. However, unlike for the more massive case discussed below, Si$_3$N$_4$, 
a rare presolar grain type, is not found among the condensates in this 1.15 \msun nova model.

\subsection {Condensates for 1.35 \msun ONe nova ejecta}

The condensate stabilities in the 1.35 \msun ONe nova model are shown in Fig. 10 (lower panel). Of the three cases investigated here, 
this is the only one with a C/O ratio above unity, in the specific shells considered. 
Condensation of graphite starts at a relatively high temperature of 1960 K. The carbides of silicon and titanium follow at 
1690 and 1660 K, respectively. When TiC starts forming in addition to SiC, graphite is no longer stable because the Si and 
C abundances are approximately the same and SiC consumes carbon. However, graphite appears again at low temperatures ($\sim 950$ K) 
when SiC is no longer stable. In addition to SiC, silicon condenses as iron silicide, silicon nitride, and silicon oxynitride. 
The latter two compounds are responsible for the termination of SiC stability and these compounds form mainly because the 
nitrogen abundance is about five times that of silicon (or carbon). For similar reasons, TiC is replaced by TiN near 1240 K. 
Moreover, aluminum nitride begins to condense at 1080 K but corundum becomes more stable at 980 K. Ca and Mg both form sulfides 
below $\sim 1000$ K and no other calcium and magnesium compounds appear because S is more abundant than both Ca and Mg combined. 
The P abundance in this ejected shell is nearly 20 times larger than that of Fe and therefore all iron from iron silicide 
enters Fe$_2$P at 1020 K.
These results suggest that condensates from massive ONe novae could be present among the known major presolar grain types 
graphite, SiC, Si$_3$N$_4$, and corundum.

\section{Discussion}

\subsection{Mean mass-averaged values versus individual shells}

In this Section we will summarize the trends found in our
 analysis of mean mass-averaged isotopic ratios (i.e., Section 2) 
and address how they 
compare with a more detailed approach based on individual shell 
variations.

\subsubsection{N and C isotopic ratios}
The most remarkable trends found in our analysis of $^{14}$N/$^{15}$N
and $^{12}$C/$^{13}$C ratios can be summarized as follows:
\begin{itemize}
\item Large dispersion in N ratios
\item In general, larger N ratios for CO models, with huge ratios of
  about $\sim 100 - 1000$ for low mass models
\item The N ratio decreases when the adopted white dwarf mass increases,
     for both CO and ONe models
\item Similar (low) C ratios for all models
\item CO models yield, in general, lower C ratios than ONe models
\end{itemize}
As can be seen in Fig. 4, most of these trends, obtained in the
framework of mean mass-averaged ratios, still hold when considering 
individual shells. Despite
the large overlap between models, we still can argue
 that the $^{14}$N/$^{15}$N ratios are dignostic
for distinguishing between CO and ONe novae, specially when taking into account
that, according to detailed stellar evolution calculations, it is likely 
that white dwarfs with masses lower than 1.05 \msun are of the CO type, 
whereas  more massive objects would be made of ONe. This fact would eliminate the overlap between the two groups in Fig. 4 with regards to N ratios. Moreover, the present plot shows no overlap
at all around $^{14}$N/$^{15}$N $\sim 50 - 1000$, which 
reinforces our claim that such large N ratios are characteristic of CO
novae.
It is hard to assess if the dependence of the N ratio on the white
dwarf mass still holds: despite of the trend seen in Fig. 4 for mass-averaged ratios, the
big overlap makes this claim questionable if one 
gives the same relative importance to all individual shells (but see 
discussion in Section 2).
Concerning C ratios, it is clear from Fig. 4, that all models are characterized
by small ratios, regardless of the nova type, the degree of mixing
and/or the mass of the compact star. This, indeed, provides a
remarkable nova signature, especially when combined with a simultaneous
low N ratio (see Fig. 1).
Moreover, a recent estimate of the $^{12}$C/$^{13}$C ratio, ranging from 
0.88 to 1.89 (Rudy et al. 2003), inferred from near-infrared 
spectrophotometry of the non-neon nova V2274 Cygni 2001 \#1, seems
to be fully compatible with the range of values shown in Fig. 4 for some
CO novae, when individual shells are taken into account. 
Furthermore, and for similar reasons as those mentioned above, it
is hard to derive any correlation between the C ratios and the nova type
when individual layers are taken into account, although Fig. 4 suggests
that larger values can be reached in ONe novae. It is worth noting that
the largest dispersions in the $^{12}$C/$^{13}$C ratios are found in
1.35 \msun ONe models. 

\subsubsection{O and Ne ratios}
The most remarkable trends found for $^{16}$O/$^{17,18}$O 
and $^{20}$Ne/$^{21,22}$Ne are:
\begin{itemize}
\item Moderate to large $^{16}$O/$^{18}$O ratios and moderate
      $^{16}$O/$^{17}$O ratios for CO models
\item Lower O ratios found, in general, for ONe models
\item In CO models, the O ratios decrease with increasing adopted white 
       dwarf masses 
\item Similar $^{20}$Ne/$^{21}$Ne ratios in CO and ONe models, with
     an overlapping region at $\sim$ 2500 - 10,000 
\item Much larger $^{20}$Ne/$^{22}$Ne ratios in ONe models than in 
      CO models
\end{itemize}
As shown in Fig. 5, there is only a small overlap in the O ratios
between CO and ONe models
even when variation bars are considered. In fact, CO models are somewhat
concentrated towards the upper right part of the diagram, whereas ONe
models tend to cluster around the lower left corner. 
We can thus claim that CO models are characterized by larger
$^{16}$O/$^{17}$O ratios (with only a small overlap with ONe models
around $\sim 10 -20$)
whereas a much larger overlapping region, around
$\sim 30-700$, is found for $^{16}$O/$^{18}$O between the two nova types. However,
the largest  and lowest $^{16}$O/$^{18}$O ratios are still achieved  in
CO and ONe models, respectively.
The correlation found between the O ratios and the white
dwarf mass for CO models still holds, in general, 
for $^{16}$O/$^{17}$O. However,
a correlation is less pronounced for $^{16}$O/$^{18}$O
although models with 0.6, 0.8 and 1.0 CO \msun show some trend.
It is worth noting that, in general, larger variations are found for
$^{16}$O/$^{18}$O  than for $^{16}$O/$^{17}$O ratios.

Similar conclusions can be made for Ne ratios
when considering individual shells (Fig. 6),
 in particular on the remarkable differences in $^{20}$Ne/$^{22}$Ne between
 CO and ONe novae 
(CO models do not show significant variations relative to the mean
 mass-averaged values, whereas some variations are present for 
 $^{20}$Ne/$^{21}$Ne) but a larger overlap for $^{20}$Ne/$^{21}$Ne.  

\subsubsection{Al and Mg ratios}
With respect to the $^{26}$Al/$^{27}$Al, $^{24}$Mg/$^{25}$Mg, and 
$^{26}$Mg/$^{25}$Mg ratios, the main trends can be summarized
as follows:
\begin{itemize}
\item High Al ratios that overlap completely for both CO and ONe novae
\item A tendency for larger $^{26}$Al/$^{27}$Al ratios to 
     be found in nova explosions hosting more massive white dwarfs
\item Low Mg ratios, in general, for both CO and ONe novae
\item The Mg ratios are nearly independent of the white dwarf mass for
      ONe models, whereas they show a complicated dependence pattern 
      in CO novae
\end{itemize}
In general, moderate dispersions are obtained for both Mg and Al isotopic 
ratios. Complete overlap characterizes the Al ratio plot (see
Fig. 7). While high Al ratios are
found for both CO and ONe novae (providing another characteristic signature
of nova outbursts), no clear dependence on the white dwarf mass is found.
However, the maximum $^{26}$Al/$^{27}$Al ratios are obtained for
the maximum white dwarf masses adopted for both CO and ONe populations.
The extraordinary overlap shown in Fig. 8
does not allow us to discern any trend in the data for both CO and ONe models. 

\subsubsection{Si ratios}
Our final analysis involves the Si isotopic ratios,; they are characterized by
the following trends in our models of classical novae:
\begin{itemize}
\item Close-to or lower-than-solar $^{29}$Si/$^{28}$Si 
      and close-to-solar $^{30}$Si/$^{28}$Si ratios for all 
       CO models
\item Close-to or lower-than-solar $^{29}$Si/$^{28}$Si ratios 
      in  all ONe models
\item Close-to or lower-than-solar $\delta ^{30}$Si/$^{28}$Si ratios for
      ONe models with $M_{wd} \leq 1.15$ \msun, but large 
      $^{30}$Si excesses for $M_{wd} \geq 1.25$ 
      ONe \msun 
\end{itemize}
 In general, all models are characterized by very small dispersions among individual shells
 (a remarkable exception being the $^{30}$Si excesses in the 1.35 \msun
ONe models. See Fig. 9). This fact reinforces most of our conclusions based
on mean mass-averaged ratios: for instance, CO models with $M_{wd} \leq 1.0$
\msun are essentially characterized by close-to-solar Si isotopic ratios (i.e.,
$\delta ^{29,30}$Si/$^{28}$Si $\sim 0$), whereas CO models of 1.15 \msun
exhibit a noticeable lower-than-solar $^{29}$Si/$^{28}$Si ratio.
Figure 9 suggests also that the dependence of the $^{30}$Si excesses 
on the white dwarf mass for ONe models holds for individual shells: as 
the mass of the white dwarf is
increased we move from lower-than or close-to-solar 
$^{30}$Si/$^{28}$Si ratios (i.e., $M_{wd} \leq 1.15$ \msun) to
a region characterized by moderate to huge $^{30}$Si excesses. Notice
that the huge excess found in all 1.35 \msun models provides a valuable and 
characteristic isotopic signature of a classical nova outburst on a
massive ONe white dwarf. 
A final, interesting aspect, concerns the $^{29}$Si/$^{28}$Si ratios
found in ONe models. As shown in Fig. 9, this ratio seems to increase as the
mass of the white dwarf increases. This trend, while evident when considering
only mean mass-averaged quantities, seems less robust for individual
shells, because of a tiny overlap between models. A remarkable situation is 
found for 1.35 \msun ONe models, which exhibit an extraordinary dispersion in 
 $\delta ^{29}$Si/$^{28}$Si: the analysis reveals that whereas a  
 large number of ejected shells are characterized by lower-than and
 close-to-solar ratios, a few shells show huge $^{29}$Si excesses. 
 In principle, this may open up the possibility to form 
grains with excesses in both $^{29}$Si and $^{30}$Si in specific shells.

\subsection{The effect of nuclear uncertainties}

Nuclear uncertainties associated with specific reaction rates important for nova nucleosynthesis may affect, to
some extent,  
the predicted isotopic ratios for a number of elements. In many cases, estimates of the impact of such nuclear 
uncertainties are obtained from post-processing calculations with temperature and density profiles that, in the
best cases, are taken from 
hydrodynamic models. Usually, such an approach has to be taken with caution, since the lack of convective 
mixing in these parametrized calculations tends to overestimate the influence of a given nuclear uncertainty.

According to a recent analysis of the effect of uncertainties in nuclear
 reaction rates for nova nucleosynthesis (Iliadis et al. 2002), present reaction rate estimates give reliable 
predictions for both $^{12}$C/$^{13}$C and $^{14}$N/$^{15}$N isotopic abundance ratios, in agreement with 
several tests performed with hydrodynamic models of nova outbursts. In contrast, uncertainties in several reactions 
can introduce large variations in the final yields for a number of species between Ne and Si. The reader is referred 
to Iliadis et al. (2002) for a complete list of reactions whose uncertainties may affect predictions for 
a number of isotopes in the framework of nova nucleosynthesis. However, because of the parametric approach adopted 
in that paper, the impact of each individual reaction has to be tested properly with a full hydrodynamic calculation.

Reactions whose impact has been confirmed through a series of hydrodynamic tests include $^{17}$O(p,$\alpha$)$^{14}$N, 
$^{17}$O(p,$\gamma$)$^{18}$F, and $^{18}$F(p,$\alpha$)$^{15}$O, which may significantly affect the $^{17,18}$O yields, 
$^{21}$Na(p,$\gamma$)$^{22}$Mg, $^{22}$Na(p,$\gamma$)$^{23}$Mg, and to some extent $^{22}$Ne(p,$\gamma$)$^{23}$Na, 
which may affect $^{21,22}$Ne ($^{22}$Na), and $^{30}$P(p,$\gamma$)$^{31}$S. 
Recent experiments focussed on 
$^{18}$F(p,$\alpha$)$^{15}$O (Bardayan et al. 2002; S\'er\'eville et al. 2003)
$^{21}$Na(p,$\gamma$)$^{22}$Mg (Bishop et al. 2003; Davids et al. 2003),
 $^{22}$Na(p,$\gamma$)$^{23}$Mg (Jenkins et al. 2004), and 
 $^{30}$P(p,$\gamma$)$^{31}$S (Rehm \& Lister, 2003) 
 have substantially improved this issue.
Indeed, a dramatic example is provided by $^{30}$P(p,$\gamma$):  
the $\delta ^{29}$Si/$^{28}$Si values are substantially reduced when the upper limit for the $^{30}$P(p,$\gamma$) rate, 
instead of the nominal one, is adopted (see Jos\'e et al. 2001, for details). 
Moreover, 
larger differences are found for 
$\delta ^{30}$Si/$^{28}$Si: the $^{30}$Si excesses obtained with the nominal rate increase by up to a factor 
$\sim 6$ if the lower limit is adopted, or even turn into deficits with the 
upper limit. Clearly, a better 
determination of this critical rate is needed in order to provide more robust 
predictions for the Si isotopic ratios.

\subsection{Comparison with other calculations}

We compared our theoretical predictions with results obtained from similar hydrodynamic models of nova outbursts by 
Kovetz \& Prialnik (1997) and Starrfield et al. (1997) for CO novae, and by Starrfield et al. (1998) for ONe 
novae. It is worth mentioning that only mean-mass averaged ratios have been considered
in the abovementioned papers, consequently our comparison will be restricted to this 
particular approach.
In general, there is good agreement with the calculations reported by Kovetz \& Prialnik (1997) and by 
Starrfield et al. (1998) for novae hosting CO white dwarf cores, in particular for $^{12}$C/$^{13}$C and $^{16}$O/$^{17}$O 
ratios. One difference involves the range of $^{14}$N/$^{15}$N ratios predicted for nova outbursts. The very high 
$^{14}$N/$^{15}$N ratios reported by Kovetz \& Prialnik (1997) and Starrfield et al. (1997) are obtained in 
explosions that achieve low peak temperatures (i.e. involve low-mass white dwarfs), for which 
$^{14}$N(p,$\gamma$)$^{15}$O($\beta^+$)$^{15}$N is not very efficient, thus reducing the $^{15}$N content and 
increasing the final $^{14}$N/$^{15}$N ratio. This interpretation is fully consistent with the results presented 
in this paper for the $0.6$ \msun CO white dwarf model, which achieves the highest N ratio. Other differences may
result from the specific reaction rate libraries adopted, from details of the treatment of convective transport, or 
from additional input physics. 

Concerning ONe models, there is also an excellent agreement with the calculations reported by Starrfield et al. 
(1998) for many isotopic ratios, including $^{12}$C/$^{13}$C, $^{26}$Al/$^{27}$Al, and $\delta^{29,30}$Si/$^{28}$Si. 
We stress that, besides the expected differences attributable to the specific choice of input physics, as mentioned
above, the main source 
of differences is probably the specific prescription adopted for the initial amounts of O, Ne, and Mg in 
the outer shells of the white dwarf, where mixing with the solar-like accreted material takes place. Whereas 
calculations by Starrfield et al. (1998) assume a core composition based on hydrostatic models of carbon-burning 
nucleosynthesis by Arnett \& Truran (1969), rather enriched in $^{24}$Mg (with ratios 
$^{16}$O:$^{20}$Ne:$^{24}$Mg $\sim 1.5:2.5:1$), we use a more recent prescription, taken from stellar evolution 
calculations of intermediate-mass stars (Ritossa et al. 1996), for which the $^{24}$Mg content is much lower 
($^{16}$O:$^{20}$Ne:$^{24}$Mg = 10:6:1). It is worth mentioning that calculations based on the Arnett \& Truran (1969)
abundances yield an unrealistically high contribution of novae to
the Galactic $^{26}$Al content, in contradiction with the 
results derived from the COMPTEL map of the 1809 keV $^{26}$Al emission in the Galaxy (see Diehl et al. 1995), 
which points towards young progenitors (type II supernovae and Wolf-Rayet stars). 
For the purpose of comparison, we list in Tables 2 \& 3 model ONeMg1, for which we assumed a
 1.25 \msun ONeMg WD, with the chemical abundances given by Arnett \& Truran (1969). As expected, this model 
agrees much better with the chemical patterns of the ejecta in the series of models of nova outbursts reported
by Starrfield et al. 

\subsection{The formation of C-rich dust in CO novae}

 The equilibrium condensation sequences reported in Section 3 predict, for the first time,
 the types of grains that can be expected to form in the ejecta of both CO and ONe novae. This includes some 
 contribution to the major presolar grain types, namely corundum (CO \& ONe
 novae), silicon carbide (ONe novae) and silicon nitride (only in massive ONe novae).
 These results confirm that SiC grains are likely to condense in ONe novae, giving support to the inferred
 ONe nova origin to presolar SiC and graphite grains recently discovered in the Murchison and
 Acfer 094 meteorites (Amari et al. 2001; Amari 2002).
  Indeed, silicon carbide and/or carbon dust formation
has been inferred through infarred measurements in a number of ONe novae, such as Nova Aql 1982 or Nova Her 1991
(see details in Gehrz et al. 1998).
 Nevertheless, it is important to point out that
 we may be facing a problem of limited statistics so conclusions exclusively based on the experimental
 determinations for only 7 grains can induce a clear bias in our global picture of classical nova outbursts.
 It is our hope that the recent implementation of new devices, such as the
 NanoSIMS (Stadermann, Walker, \& Zinner 1999a, 1999b; Hoppe 2002), will improve soon the statistics and
 will help us to extract conclusions in a firmer basis, providing in turn a tool to constraint theoretical 
 nova models. We note that three additional nova candidate grains have recently been located (Nittler \& Hoppe 2004). 

A puzzling preliminary result obtained in our analysis of equilibrium condensation sequences is that 
reduced condensates such as SiC or graphite do not form in CO novae (at least for the
selected 1.15 \msun CO case), and hence, they will not contribute to presolar SiC or graphite grains. Whereas a much
deeper analysis of ejecta from a wider sample of CO nova models is required to confirm this result, it remains
to be understood which mechanism is responsible for the formation of C-rich dust seen in infrared analyses of 
CO novae, a feature that seems to be
common in many explosions of this type (see Gehrz et al. 1998). Possible explanations include a mechanism capable of
dissociating the CO molecule (see Clayton, Liu \& Dalgarno 1999, for a radiation-based mechanism to dissociate the CO 
molecule in a 
supernova environment), that would drive the condensation sequence out of equilibrium conditions
(however, some aspects of the chemistry where the CO molecule is absent have been
investigated by Ebel \& Grossman (2001), showing that SiC formation is still unlikely).
In this respect, the recent
spectrophotometric studies of CO emission in nova V2274 Cygni 2001 \#1 (Rudy et al. 2003) at two different 
epochs suggest 
that, whereas emission from the first overtone of carbon monoxide is seen about 18 days after outburst, the absence of 
such CO emission at 370 days is an indication of partial destruction of the CO molecules. 
Among the mechanisms proposed are photodissociation
and photoionization (see Shore \& Gehrz 2004), charge transfer reactions 
(Rawlings 1988; Liu, Dalgarno, \& Lepp 1992) and dissociation by He$^+$ ions (Lepp, Dalgarno, \& McCray 1990). 
In addition, Scott (2000) has suggested that 
rotation-driven latitudinal abundance gradients may affect dust formation. 
Other alternatives involve possible contamination of the outer layers of the main sequence
companion (during the previous evolution of the white dwarf progenitor), that in some case may
lead to C-enrichment in those shells, or 
scenarios leading to nova
explosions with significant C-enriched envelopes, that may lead to C$>$O ejecta. 
In this respect, we have performed a hydrodynamic
simulation of a 0.6 \msun CO white dwarf, identical to the model previously discussed in this paper but with a slightly
different composition for the outermost layers of the white dwarf core (for which 60\% $^{12}$C and 
40\% $^{16}$O has been adopted instead of the usual $^{12}$C/$^{16}$O=1). 
The results of this test suggest that indeed the outermost ejected envelope is C-rich,
allowing for the formation of C-rich dust. Finally, the recent update of the solar
abundances (see Lodders 2003), that reduce the C and O content in the solar mixture 
by about $\sim$ 50\%,
may help to condense C-rich dust in CO novae thanks to the presence of Si, Mg and Al
atoms (in a similar way as described for ONe novae). Hydrodynamic tests to validate this 
possibility are currently under way.

\section{Conclusions}

In this paper, we presented a detailed analysis of isotopic ratios in the ejecta of classical novae,
for nuclei up to Si, based on a series of 20 hydrodynamic models of the 
explosion. Both analysis based on global mean mass-averaged ratios and 
on composition gradients through individual shells were presented.
From this study, we conclude that nova grains are, in general, 
characterized by low C ratios, high Al ratios, and close-to- or
slightly lower-than-solar $^{29}$Si/$^{28}$Si ratios. Other predicted 
isotopic ratios are specific of
each nova type (CO or ONe): for instance, we expect that grains condensed in the ejecta from massive ONe
novae will exhibit significant $^{30}$Si excesses (with the posibility of
a $^{29}$Si excess not being ruled out in the outermost ejected shells), 
whereas those resulting from explosions in CO novae will
show close-to-solar $^{30}$Si/$^{28}$Si ratios. Indeed, our study suggests
that the ejecta from ONe novae are characterized by low N ratios whereas CO 
novae show a large dispersion in the N ratios, with values ranging from 
$\sim$0.1 to more than 1000. With respect to Ne, ONe novae are characterized by large $^{20}$Ne/$^{21}$Ne and
$^{20}$Ne/$^{22}$Ne ratios, whereas CO novae show large $^{20}$Ne/$^{21}$Ne but small $^{20}$Ne/$^{22}$Ne
ratios.  However, it is worth noting that predictions of N and Ne isotopic ratios in the grains are difficult 
because of a likely N isotopic equilibration, and also because $^{22}$Ne excesses could come both from $^{22}$Ne 
implantation or from $^{22}$Na in situ decay. 

In addition, we report on equilibrium condensation sequences that predict, for
the first time, the types of grains that are expected to form in the ejecta of both CO and ONe novae. 
 Our analysis shows that the ejecta of 1.15 \msun CO novae are likely contributors to the known presolar 
 populations of corundum, spinel, and enstatite grains. 1.15 \msun ONe novae can
  produce corundum and enstatite grains as well as SiC grains. The more
  massive 1.35 \msun novae allow formation of corundum, silicon carbide, and
  silicon nitride grains.
 This analysis points out that SiC grains are likely to condense in the ejecta from ONe novae
 and supports the inferred ONe nova origin of the sample of presolar SiC and graphite grains isolated 
 from the Murchison and Acfer 094 meteorites (Amari et al. 2001; Amari 2002).
Among the presolar oxide grains discovered so far no oxide grain with a nova signature has been
discovered to date, although they are likely to condense in most (if not all) nova explosions, according to this work. 
These grains would be clearly identified by huge $^{17}$O and somewhat smaller $^{18}$O excesses. 

We expect that these theoretical estimates will help to correctly identify nova grains embedded in primitive
meteorites. 
Indeed, the recent development of new instruments, such as the NanoSIMS is expected to lead
to future identification of nova grains. The improved spatial resolution and
sensitivity of this instrument, together with its capability to measure simultaneously several isotopes,
opens new possibilities, including measurements of elemental and isotopic compositions
inside the grains (see Stadermann et al. 2002). Such accurate sources of information will help to constrain
nova models in a much more precise way. 

\section{Appendix. Nucleosynthesis and Nova Models}

Tables 2 \& 3 list the mean mass-averaged isotopic ratios in the ejected 
envelopes from a sample of 20 hydrodynamic models of classical nova 
outbursts (Jos\'e \& Hernanz 1998; 
Hernanz et al. 1999; Jos\'e et al. 2001, and unpublished data).
Calculations have been carried out by means 
of the one-dimensional, implicit, Lagrangian, hydrodynamical code {\it SHIVA}. 
Minimum and maximum ratios for individual ejected shells (in square brackets) 
are also given for completeness. 

Each model listed in Tables 2 \& 3 is characterized by the mass of the 
underlying white dwarf, as well as by the initial envelope composition, which 
distinguishes explosions taking place on 
 white dwarfs hosting either CO or ONe cores. As discussed in Jos\'e \& Hernanz (1998), the models assume mixing 
between material from the outermost core and the solar-like accreted envelope 
(see also Starrfield el al. 1998, and references 
therein) in order to mimic the unknown mechanism responsible for the 
enhancement in metals, essentially $^{12}$C, which ultimately powers 
the explosion. 
To parametrize this process, different degrees of mixing, ranging from 25\% to 
75\%, have been considered and are also indicated in the Tables.
The adopted composition of the outer layers for CO white dwarfs is 
X($^{12}$C)=0.495, X($^{16}$O)=0.495, 
and X($^{22}$Ne)=0.01 (Salaris et al. 1997).
For ONe white dwarfs we used X($^{16}$O)=0.511, X($^{20}$Ne)=0.313, X($^{12}$C)=$9.16 \times 10^{-3}$, 
X($^{23}$Na)=$6.44 \times 10^{-2}$, X($^{24}$Mg)=$5.48 \times 10^{-2}$, X($^{25}$Mg)=$1.58 \times 10^{-2}$, 
X($^{27}$Al)=$1.08 \times 10^{-2}$, X($^{26}$Mg)=$9.89 \times 10^{-3}$, X($^{21}$Ne)=$5.98 \times 10^{-3}$, and 
X($^{22}$Ne)=$4.31 \times 10^{-3}$. 
These values correspond to the composition of the remnant of a
10 \msun population I star, evolved from the H-burning main sequence phase up to the thermally pulsing 
super-asymptotic giant branch stage (Ritossa et al.  1996). Solar abundances were taken from Anders \& Grevesse (1989).
For comparison, we include also results from model ONeMg1, where the adopted
chemical composition of the white dwarf core is taken from carbon-burning nucleosynthesis calculations by 
Arnett \& Truran (1969) (see Section 4.3, for details). 
We note that this rather old prescription for the white dwarf is the one assumed in all 
calculations of ONe(Mg) novae published by Starrfield's group up to now. 

Table 3 summarizes the different initial isotopic ratios in the
sample of models presented in this paper. 
Although the initial $^{13}$C abundance is the same in both CO and ONe novae 
(i.e., a fraction of the solar content, depending on the adopted degree of mixing), the much higher initial $^{12}$C content 
in CO models results in initial $^{12}$C/$^{13}$C ratios different from those in ONe models: they range from $\sim$ 5000 
to 44,000 (25\% to 75\% mixing) for CO, and from $\sim$ 180 to 900 
(25\% to 75\% mixing) for ONe models.  The initial $^{14}$N/$^{15}$N ratio for 
all models is solar.
 Because oxygen in the white dwarf (for both CO and ONe models) is almost pure $^{16}$O, the initial oxygen ratios are extremely high. 
Values depend on the degree of mixing, which strongly modifies the $^{16}$O content in the envelope: for CO models, 
the initial $^{16}$O/$^{18}$O ratios range from 9100 (25\% mixing) to 78,000 (75\% mixing), whereas $^{16}$O/$^{17}$O ratios
range from 48,000 (25\%) to 410,000 (75\%). Similar values are found for ONe models: $^{16}$O/$^{18}$O ratios range 
from 9500 (25\%) to 80,000 (75\%), whereas $^{16}$O/$^{17}$O ratios range from 49,000 (25\%) to 420,000 (75\%).
The $^{20}$Ne/$^{22}$Ne initial ratios are larger in ONe novae
(i.e., 74 - 78, depending on the degree of mixing)
than in CO novae (0.06 for 75\% mixing to 0.5 for 25\% mixing). The initial 
$^{20}$Ne/$^{21}$Ne ratio is 412 (the solar value) for the CO models, whereas a
value around 55 corresponds to the ONe models. 
CO models have initially solar Mg ratios (i.e., $^{26}$Mg/$^{25}$Mg= 1.1; $^{24}$Mg/$^{25}$Mg=7.9), 
since Mg is only present in the accreted material. 
In contrast, ONe models are characterized by  
$^{26}$Mg/$^{25}$Mg= 0.6 and $^{24}$Mg/$^{25}$Mg=3.6 (where Mg from both core 
material and accreted envelope is taken into account). In both cases, the 
initial isotopic ratios are nearly independent of the adopted degree of mixing. 
Finally, all silicon initially present in the envelope comes from the white 
dwarf companion in solar proportions 
(i.e., $\delta ^{29,30}$Si/$^{28}$Si = 0), regardless of the nova type and of 
the degree of mixing. 

Most CO and ONe models listed in Tables 2 \& 3 have been computed with the 
same nuclear reaction network, consisting of $\sim 100$ isotopes, ranging 
from $^1$H to $^{40}$Ca and linked through a net containing
370 nuclear reactions (details can be found in Jos\'e \& Hernanz 1998).
Exceptions are models ONe4, ONe7, CO2 and CO8, for which updated
$^{18}$F+p rates have been used (see Hernanz et al. 1999, for details), 
and models CO1 and ONe9, for which both $^{18}$F+p and S-Ca updated rates
have been taken into account (see Jos\'e et al. 2001, for details).

\acknowledgments

We acknowledge the efforts of an anonymous referee who helped improving the manuscript.
We are grateful to D. Clayton, A. Davis, E. Garc\'\i a--Berro, J. Guerrero,
and J. Truran for stimulating discussions 
and suggestions on several topics addressed in this manuscript. This work has been partially supported by the MCYT grants
AYA2001-2360 and AYA2002-0494C03-01 (JJ and MH), 
NASA grants NAG5-4545 (SA and EZ) and  NAG5-10553 (KL),
and by the E.U. FEDER funds. JJ acknowledges support from the catalan AGAUR
during a sabbatical leave.

\clearpage 
\begin{deluxetable}{cccccccc}
\tabletypesize{\scriptsize}
\tablecaption{Presolar grains with an inferred nova origin.}
\tablewidth{0pt}
\tablehead{
\colhead{Grain\tablenotemark{a}}&
\colhead{Composition}&
\colhead{$^{12}$C/$^{13}$C}&
\colhead{$^{14}$N/$^{15}$N}&
\colhead{$\delta$($^{29}$Si/$^{28}$Si)}&
\colhead{$\delta$($^{30}$Si/$^{28}$Si)}&
\colhead{$^{26}$Al/$^{27}$Al}&
\colhead{$^{20}$Ne/$^{22}$Ne}
}
\startdata
AF15bB-429-3 (Ama01) & SiC &$9.4 \pm 0.2$ &    &$28 \pm 30$  &$1118 \pm 44$  & &\\
AF15bC-126-3 (Ama01) & SiC &$6.8 \pm 0.2$ &$5.22 \pm 0.11$ &$-105 \pm 17$  &$237 \pm 20$  & &\\
KJGM4C-100-3 (Ama01) & SiC &$5.1 \pm 0.1$ &$19.7 \pm 0.3 $ &$55 \pm 5$  &$119 \pm 6$  & 0.0114 &\\
KJGM4C-311-6 (Ama01) & SiC &$8.4 \pm 0.1$ &$13.7 \pm 0.1 $ &$-4 \pm 5$  &$149 \pm 6$  & $>$0.08 &\\
KJC112 (Hop95)  & SiC &$4.0 \pm 0.2$ &$6.7 \pm 0.3$ &     &     & & \\
KFC1a-551 (Ama01) & C &$8.5 \pm 0.1$ &$273 \pm 8$  &$84 \pm 54$  &$761 \pm 72$  & & \\
KFB1a-161 (Nic03) & C &$3.8 \pm 0.1$ &$312 \pm 43$ &$-133 \pm 81$&$37 \pm 87$  & & $ < 0.01$ \\
\hline
Solar       & & 89   & 272   &     &   & & 14 \\
Nova Models      & & $0.3 - 1.8$ & $0.3 - 1400$ & $(-900) - 10$ & $(-1000) - 47000$&
           $0.01 - 0.6$ & $0.1 - 250$\\
\enddata
 \tablenotetext{a}{The Solar N ratio in the Table is that of the air. Grains AF... are from the Acfer 094 meteorite,
  whereas grains KJ... and KF... are from the Murchison meteorite. 
  Delta values measure deviations from the solar Si isotopic ratios in permil
  (see Section 2.4 for definition.). Errors are 1$\sigma$. }
\end{deluxetable}

\clearpage

\begin{deluxetable}{ccccccccc}
\tabletypesize{\scriptsize}
\rotate
\tablecaption{Isotopic ratios for C, N, O, and Ne, obtained from different 
models, as displayed in Figs. 4 to 6. }
\tablewidth{0 pt}
\tablehead{
\colhead{Model\tablenotemark{a}}&
\colhead{Mass (\msun)}&
\colhead{Initial composition}&
\colhead{$^{12}$C/$^{13}$C}&
\colhead{$^{14}$N/$^{15}$N}&
\colhead{$^{16}$O/$^{17}$O}&
\colhead{$^{16}$O/$^{18}$O}&
\colhead{$^{20}$Ne/$^{21}$Ne}&
\colhead{$^{20}$Ne/$^{22}$Ne}
}
\startdata 
 ONe1  & 1.00 (JH98) & 50\% ONe  & 0.83 [0.59-1.4]& 3.6  [0.59-27.8]& 10  [7.1-17.7]& 23.3 [15.3-73.6]& 
                                   9950 [4320-27,400] &  97  [89.9-103]\\
 ONe2  & 1.15 (JH98) & 25\% ONe  & 0.85 [0.67-1.9]& 1.9  [0.55-3.6] & 2.3 [1.5-4.2] & 11.3 [8.9-23.6] &
                                   7270 [3990-9170]   & 159  [121-169]  \\
 ONe3& 1.15 (JH98) & 50\% ONe    & 0.89 [0.23-1.6]& 1    [0.39-2.4] & 4.6 [3.3-18.5]&22.5 [10.5-63.8] & 
                                   6300 [3920-8290]& 113  [100-120]   \\
 ONe4&1.15 (HJCGI99)&50\% ONe    & 0.76 [0.63-2.0]& 1.2  [0.36-1.9] & 3.8 [2.5-5.4] & 388  [309-663]  & 
                                   2200 [966-3540]    & 109  [72.9-119] \\
 ONe5  & 1.15 (JH98) & 75\% ONe  & 0.88 [0.65-1.8]& 1.2  [0.35-3.5] & 6   [4.3-10.6]& 35.5 [22.2-183] & 
                                   6830 [4600-11,900] & 108  [78.1-118] \\
 ONe6  & 1.25 (JH98) & 50\% ONe  & 0.95 [0.73-2.2]& 0.82 [0.30-1.5] & 1.9 [1.5-3.3] & 19   [13.9-49.3]& 
                                   5400 [3510-7080] & 181  [113-190]  \\
 ONe7 & 1.25 (HJCGI99)&50\% ONe & 0.79 [0.60-2.4]& 0.97 [0.28-1.6] & 1.9 [1.4-2.9] & 220  [188-394]  &
                                   1620 [804-2770]& 153 [77.9-185]\\
 ONe8  & 1.35 (JH98) & 50\% ONe & 1.5  [0.50-3.2]& 0.41 [0.14-0.61]& 1.4 [0.63-2.6]& 25.3 [22.7-76.2]& 
                                   3090 [2460-3340]&220 [56.9-459]\\
 ONe9 & 1.35 (JCH01) & 50\% ONe & 1.5  [0.54-3.1]& 0.40 [0.14-0.58]& 1.4 [0.68-2.9]& 283  [176-529]  &
                                   557  [426-641] & 125 [37.2-352]\\
 ONe10  & 1.35 (JH98) & 75\% ONe & 1.1  [0.28-2.7]& 0.25 [0.12-0.52]& 1.9 [1.7-3.7] & 56.9 [46.6-265]& 
                                   3000 [2480-4210]&247 [79.9-1430]\\
 \hline
 ONeMg1 & 1.25   &50\% ONeMg     & 0.73 [0.52-2.4]& 1.0  [0.34-1.6] & 2.6 [2-8.9]   & 22   [11.2-211] & 
                                   5060 [4190-5740]&1550 [818-2890]\\
 \hline
 CO1 & 0.6   & 50\% CO           & 1.8  [1.7-1.8] & 1370 [742-1960] & 232 [230-236] &38,800 [33,800-43,000]& 
                                 26,750 [20,300-33,400]&0.18 [0.18-0.18]\\
 CO2 & 0.8 (HJCGI99) & 50\% CO   & 0.51 [0.42-1.8]& 60   [21.4-177] & 63  [56.1-178]& 3960  [3400-20,600]&
                                 17,670 [3120-46,200]  &0.18 [0.18-0.18]\\
 CO3  & 0.8 (JH98) & 25\% CO     &0.45 [0.39-0.63]& 103  [21.3-245] &41.9 [38.8-66] & 174   [154-436] &
                                 43,500 [12,000-89,700]&0.51 [0.51-0.53]\\
 CO4  & 0.8 (JH98) & 50\% CO     & 0.52 [0.49-1.8]& 127  [20.8-142] &60.4 [57.8-178]& 502   [474-3790]& 
                                 21,500 [3080-26,800]  &0.18 [0.18-0.18]\\
 CO5  & 1.0 (JH98) & 50\% CO     &0.30 [0.29-0.70]& 19   [6-267]    &31.9 [30.9-46.8]&123   [95.2-423]& 
                                 15,900 [5480-147,000]  &0.19 [0.18-0.19] \\
 CO6  & 1.15 (JH98) & 25\% CO    &0.71 [0.50-0.78]& 3.3  [1.6-8.5]  & 7.6 [5.8-9.7] & 22    [16.1-120]& 
                                  7740  [3640-14,200]  &0.70 [0.61-0.72] \\
 CO7 & 1.15 (JH98) & 50\% CO     &0.54 [0.50-1]   & 3.0  [2.4-5.3]  &10.6 [10.4-16.7]&62.2  [35.7-443]& 
                                  5990  [2740-15,500]  &0.22 [0.2-0.23] \\
 CO8& 1.15 (HJCGI99)&50\% CO     &0.54 [0.50-0.93]& 2.8  [2.4-13.5] &10.9 [10.3-15.7]&854   [750-3730]& 
                                  2640  [1240-10,800] & 0.22 [0.2-0.23] \\
 CO9  & 1.15 (JH98) & 75\% CO    &0.39 [0.32-0.88]& 5.2  [3.2-12]   & 19  [18.6-26.5]&103   [51.3-548]& 
                                  7580  [2690-22,700]  &0.078 [0.07-0.082]\\
 \hline
 Solar ratios&    &              & 89             & 272             & 2622          & 498             & 
                                   412                 & 14             \\
\enddata
 \tablenotetext{a}{All models have been computed with the nuclear reaction
 network described in JH98, except Models ONe4, ONe7, CO2 and CO8 (HJCGI99)
 and Models CO1 and ONe9 (JCH01). See Appendix, for details.}
\end{deluxetable}

\clearpage

\begin{deluxetable}{cccccccc}
\tabletypesize{\scriptsize}
\rotate
\tablecaption{Same as Table 2, for Mg, Al and Si isotopes, as
displayed in Figs. 7 to 9.}
\tablewidth{0 pt}
\tablehead{
\colhead{Model}&
\colhead{Mass (\msun)}&
\colhead{Initial composition}&
\colhead{$^{24}$Mg/$^{25}$Mg}&
\colhead{$^{26}$Mg/$^{25}$Mg}&
\colhead{$^{26}$Al/$^{27}$Al}&
\colhead{$\delta$($^{29}$Si/$^{28}$Si)}&
\colhead{$\delta$($^{30}$Si/$^{28}$Si)}
}
\startdata
 ONe1  & 1.00 (JH98) & 50\% ONe  & 0.026 [0.005-0.09]  & 0.090 [0.082-0.11]& 0.23 [0.19-0.32]& 
                                   -951 [(-951)-(-949)]& -965 [(-965)-(-964)]\\
 ONe2  & 1.15 (JH98) & 25\% ONe  & 0.021 [0.013-0.037] & 0.096 [0.076-0.13]& 0.25 [0.23-0.3] & 
                                   -803 [(-808)-(-784)]& 296 [(-128)-320]\\
 ONe3& 1.15 (JH98) & 50\% ONe    & 0.036 [0.004-0.076] & 0.11  [0.074-0.14]& 0.22 [0.16-0.27]& 
                                   -852 [(-853)-(-848)]& -645 [(-744)-(-624)]\\
 ONe4&1.15 (HJCGI99)&50\% ONe    & 0.10  [0.04-0.29]   & 0.16  [0.14-0.18] & 0.24 [0.22-0.27]& 
                                   -788 [(-795)-(-756)]& -105 [(-407)-(-73)]\\
 ONe5  & 1.15 (JH98) & 75\% ONe  & 0.21  [0.043-0.65]  & 0.13  [0.082-0.21]& 0.22 [0.21-0.24]& 
                                   -796 [(-808)-(-756)]& -354 [(-482)-(-325)]\\
 ONe6  & 1.25 (JH98) & 50\% ONe  & 0.087 [0.032-0.19]  & 0.11  [0.083-0.15]& 0.28 [0.25-0.30]& 
                                   -701 [(-734)-(-585)]& 1380 [722-1470]\\
 ONe7 & 1.25 (HJCGI99)&50\% ONe & 0.13  [0.05-0.33]   & 0.15  [0.13-0.17] & 0.28 [0.26-0.32]& 
                                   -655 [(-698)-(-447)]& 2020 [1230-2110]\\
 ONe8  & 1.35 (JH98) & 50\% ONe & 0.089 [0.023-0.19]  & 0.14  [0.12-0.16] & 0.42 [0.26-0.46]& 
                                   -74  [(-677)-1760]  & 7730 [7260-9700]\\
 ONe9 & 1.35 (JCH01) & 50\% ONe & 0.12  [0.026-0.27]  & 0.15  [0.14-0.17] & 0.42 [0.27-0.46]& 
                                   -54  [(-666)-1830]  & 8150 [7760-9580]\\
 ONe10 & 1.35 (JH98) & 75\% ONe  & 0.19  [0.034-0.38]  & 0.15  [0.056-0.19]& 0.35 [0.19-0.39]& 
                                   -113 [(-742)-1210]  & 7140 [7000-7770]\\
 \hline
 ONeMg1 & 1.25   &50\% ONeMg     & 0.47  [0.14-1.8]    &0.043 [0.018-0.053]& 0.44 [0.34-0.88]& 
                                   -546 [(-652)-87.8]  & 3780 [3500-3830]\\
 \hline
 CO1 & 0.6   & 50\% CO           & 4.3   [4.3-4.3]     & 0.65  [0.65-0.65] &0.006 [0.006-0.006]&
                                   -3.3 [(-3.3)-(-3.3)]& 1.9 [1.9-1.9]\\
 CO2 & 0.8 (HJCGI99)  & 50\% CO  & 0.31  [0.22-1.8]    & 0.16  [0.15-0.35] &0.095 [0.021-0.11]&
                                   -6.3 [(-6.3)-(-3.3)]& -1.2 [(-1.20)-1.9]\\
 CO3  & 0.8 (JH98) & 25\% CO     & 0.16  [0.12-0.7]    & 0.14  [0.13-0.21] & 0.18 [0.074-0.2] & 
                                   -6.4 [(-6.4)-(-2.4)]& -3.2 [(-3.2)-0.85]\\
 CO4  & 0.8 (JH98) & 50\% CO     & 0.27  [0.25-1.8]    & 0.15  [0.15-0.34] & 0.14 [0.023-0.14]& 
                                   -6.3 [(-6.3)-(-3.3)]& -1.2 [(-1.2)-1.9]\\
 CO5  & 1.0 (JH98) & 50\% CO     & 0.060 [0.025-0.31]  & 0.10  [0.099-0.14]& 0.39 [0.23-0.42] & 
                                  -33 [(-35.6)-(-18.1)]& 4.9 [(-1.9)-6.5]\\
 CO6  & 1.15 (JH98) & 25\% CO    & 0.042 [0.017-0.15]  & 0.096 [0.08-0.12] & 0.25 [0.23-0.34] & 
                                   -677 [(-697)-(-497)]& -47.1 [(-53.6)-(-21.2)]\\
 CO7 & 1.15 (JH98) & 50\% CO     & 0.12  [0.024-0.69]  & 0.094 [0.073-0.18]& 0.38 [0.36-0.43] & 
                                   -449 [(-480)-(-236)]& 29.6 [15.8-45.5]\\  
 CO8 & 1.15 (HJCGI99)&50\% CO    & 0.11  [0.028-0.61]  & 0.14  [0.12-0.2]  & 0.39 [0.35-0.43] & 
                                   -435 [(-478)-(-243)]  & 32.6 [19.8-52.1]\\  
 CO9  & 1.15 (JH98) & 75\% CO    & 0.080 [0.01-0.46]   & 0.070 [0.06-0.12] & 0.58 [0.53-0.59] & 
                                   -267 [(-296)-(-159)]& 90.1 [71.3-97.4]\\
 \hline
 Solar ratios&     &             & 7.9                 & 1.1               & 0                & 
                                     0                 & 0 \\
\enddata
\end{deluxetable}

\clearpage

\begin{deluxetable}{ccccccc}
\tabletypesize{\scriptsize}
\rotate
\tablecaption{Initial isotopic ratios for both CO and ONe models, as a function
of the degree of mixing.}
\tablewidth{0 pt}
\tablehead{
\colhead{Initial ratio\tablenotemark{a}}&
\colhead{}&
\colhead{CO Models}&
\colhead{}&
\colhead{}&
\colhead{ONe Models}&
\colhead{}
}
\startdata
                     & 25\%   & 50\%    & 75\%    & 25\%   & 50\%    & 75\% \\
 \hline
 $^{12}$C/$^{13}$C   &   4987 &  14,782 &  44,165 &    181 &     362 & 906 \\
 $^{14}$N/$^{15}$N   &    271 &     271 &     271 &    271 &     271 & 271 \\
 $^{16}$O/$^{17}$O   & 47,724 & 137,929 & 408,542 & 49,182 & 142,302 & 421,663\\
 $^{16}$O/$^{18}$O   &   9064 &  26,196 &  77,592 &   9340 &  27,026 &  80,084\\
 $^{20}$Ne/$^{21}$Ne &    412 &     412 &     412 &     56 &      55 &      55\\
 $^{20}$Ne/$^{22}$Ne &   0.51 &    0.18 &   0.059 &     74 &      78 &      79\\
 $^{24}$Mg/$^{25}$Mg &    7.9 &     7.9 &     7.9 &    3.7 &     3.6 &     3.6\\
 $^{26}$Mg/$^{25}$Mg &    1.1 &     1.1 &     1.1 &   0.61 &    0.60 &    0.60\\
\enddata
 \tablenotetext{a}{For all CO \& ONe Models, $^{26}$Al/$^{27}$Al = 
 $\delta$($^{29,30}$Si/$^{28}$Si) = 0. }
\end{deluxetable}

\clearpage
  \begin{figure}
  \plotone{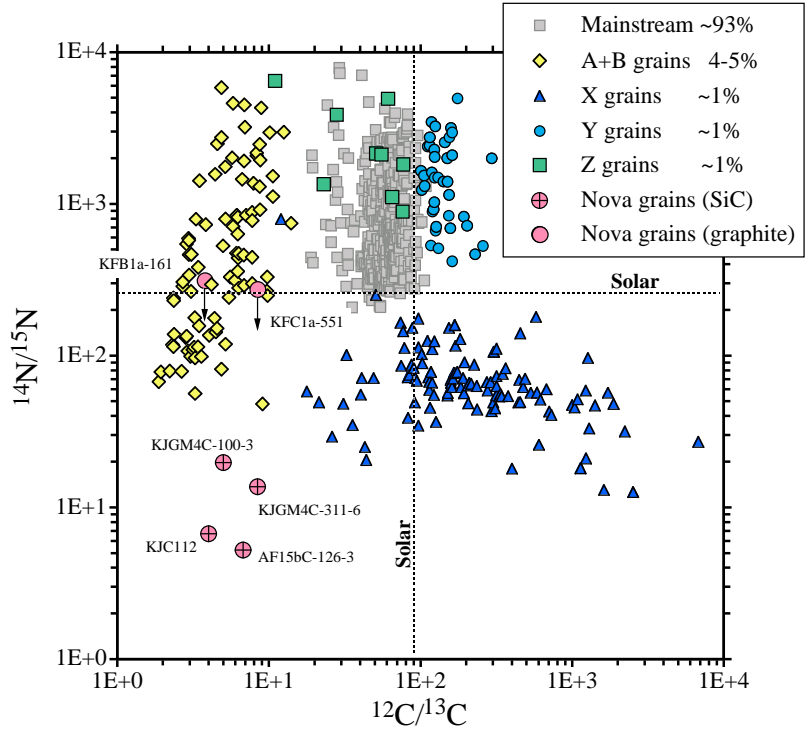}
  \caption{Carbon and nitrogen isotopic ratios of nova candidate grains are compared
  with those of SiC grains of different populations (see legend, for details). Silicon carbide grains have been
  classified into several populations based on their C, N and Si isotopic
  ratios.  Error bars are smaller than the symbols.}
  \end{figure}

  \begin{figure}
  \plotone{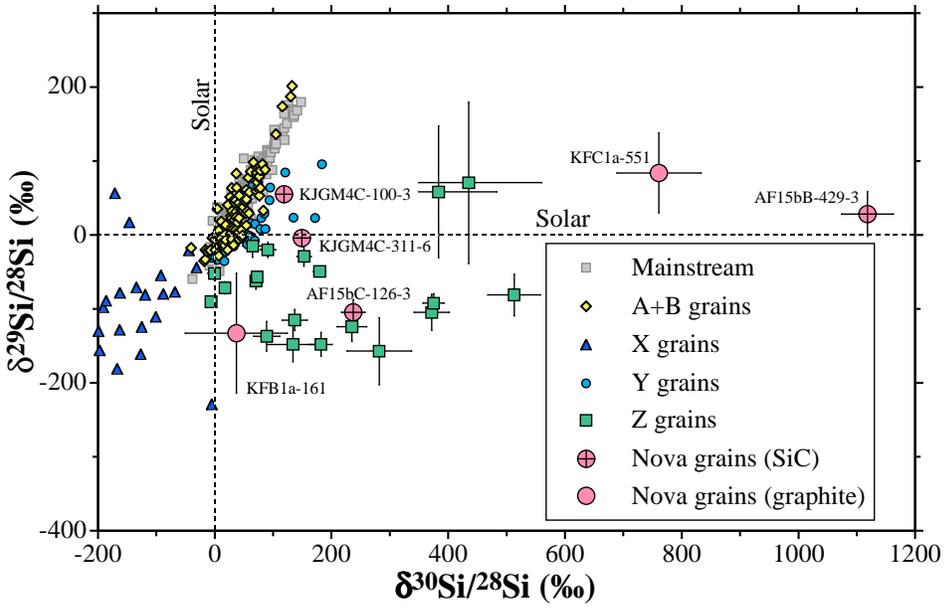}
  \caption{Silicon isotopic ratios of five nova candidate
  grains and other SiC grains. Ratios are expressed as delta values,
  deviations from the solar Si isotopic ratios in permil
  (see Section 2.4 for definition.)}
  \end{figure}

  \begin{figure}
  \plotone{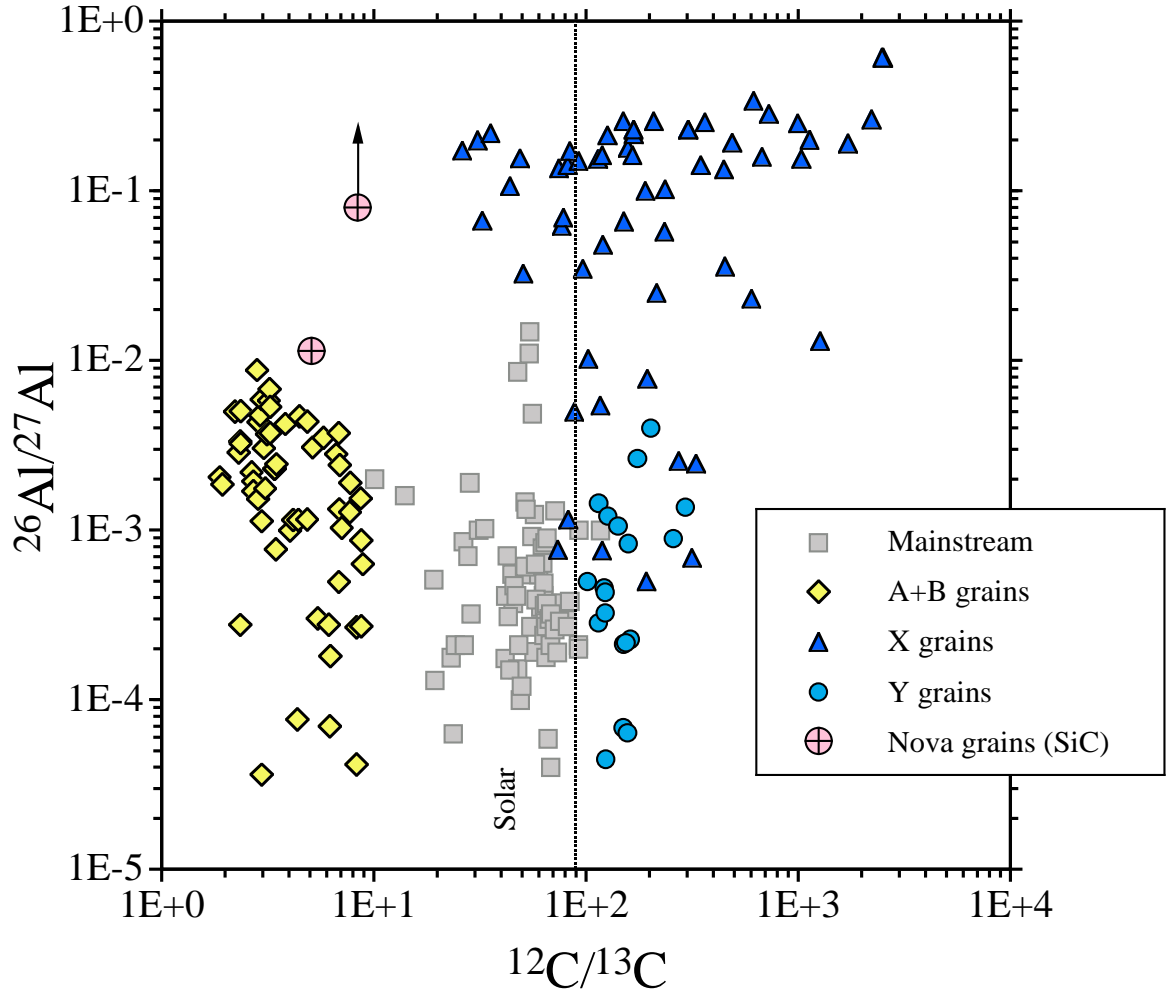}
  \caption{Same as Fig. 1, for aluminum versus carbon isotopic ratios. }
  \end{figure}

  \begin{figure}
  \plotone{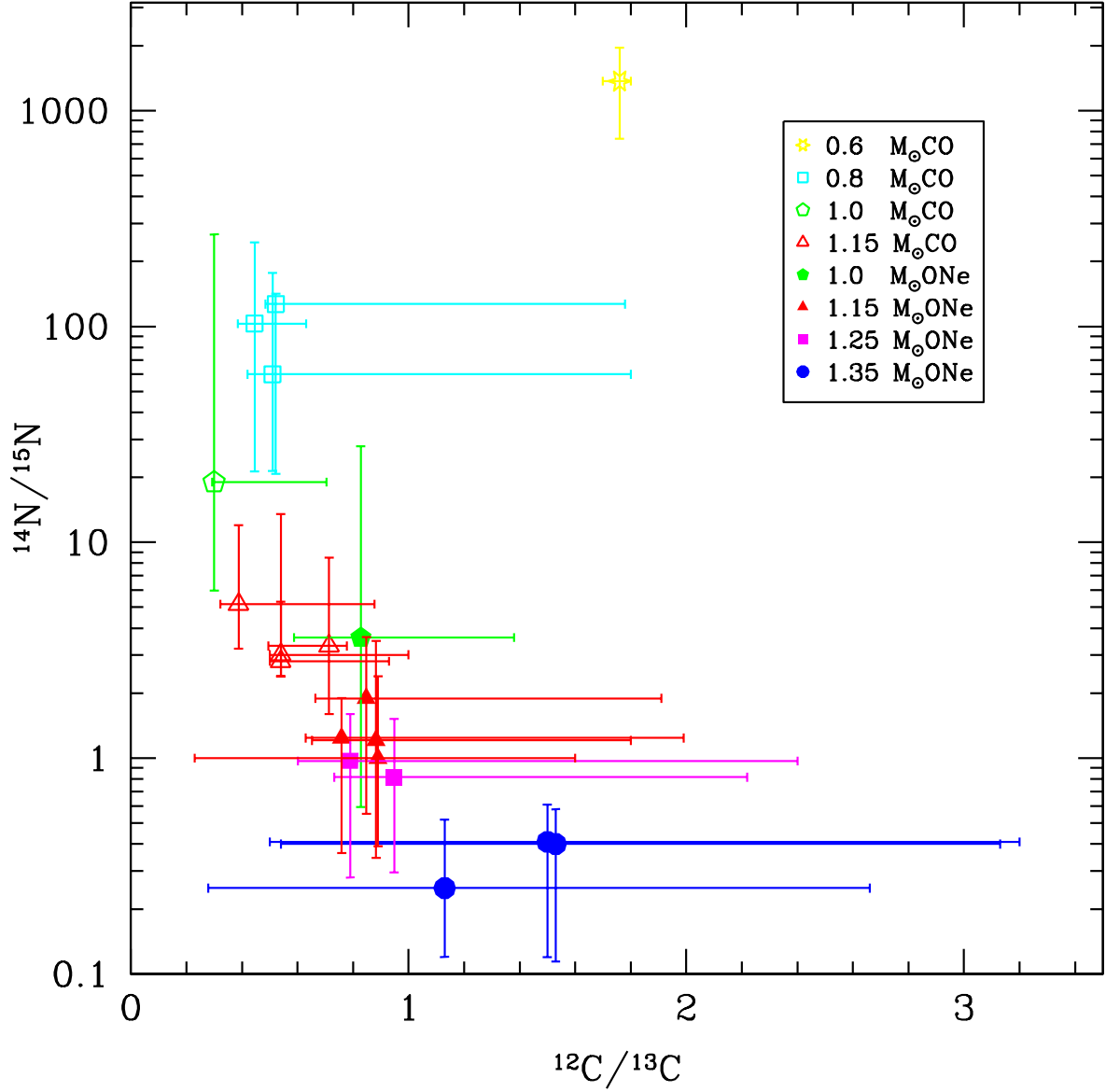}
  \caption{ Nitrogen versus carbon isotopic ratios, predicted by hydrodynamic
  models for both CO and ONe novae (see Tables 2 \& 3, for details).
  Points represent mean mass-averaged ratios. Deviation bars, taking into
  account the gradient of composition in the ejected shells, are also shown
  for all models. See text for details.}
  \end{figure}

  \begin{figure}
  \plotone{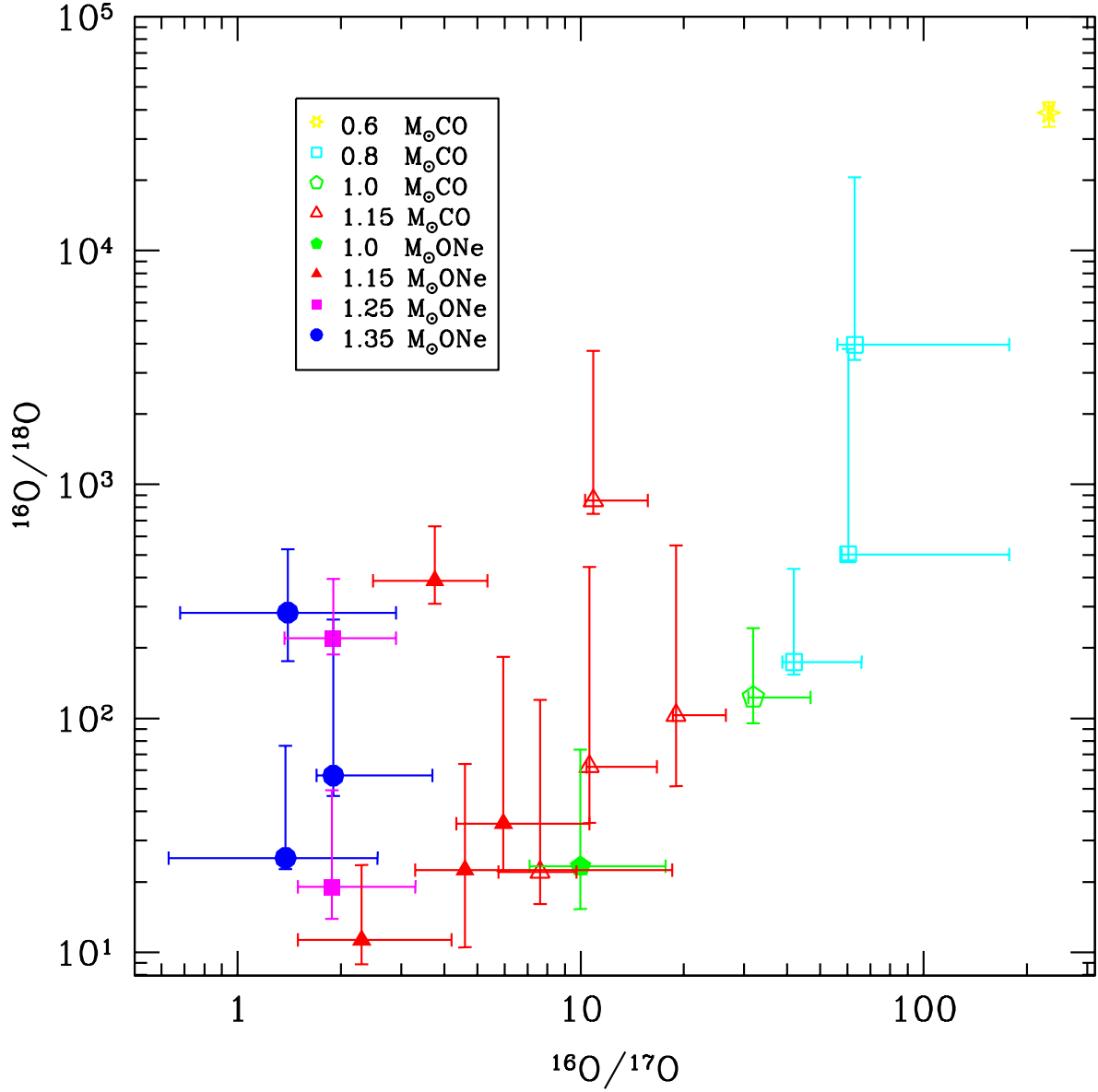}
  \caption{Same as Fig. 4, for $^{16}$O/$^{18}$O and $^{16}$O/$^{17}$O ratios. }
  \end{figure}

  \begin{figure}
  \plotone{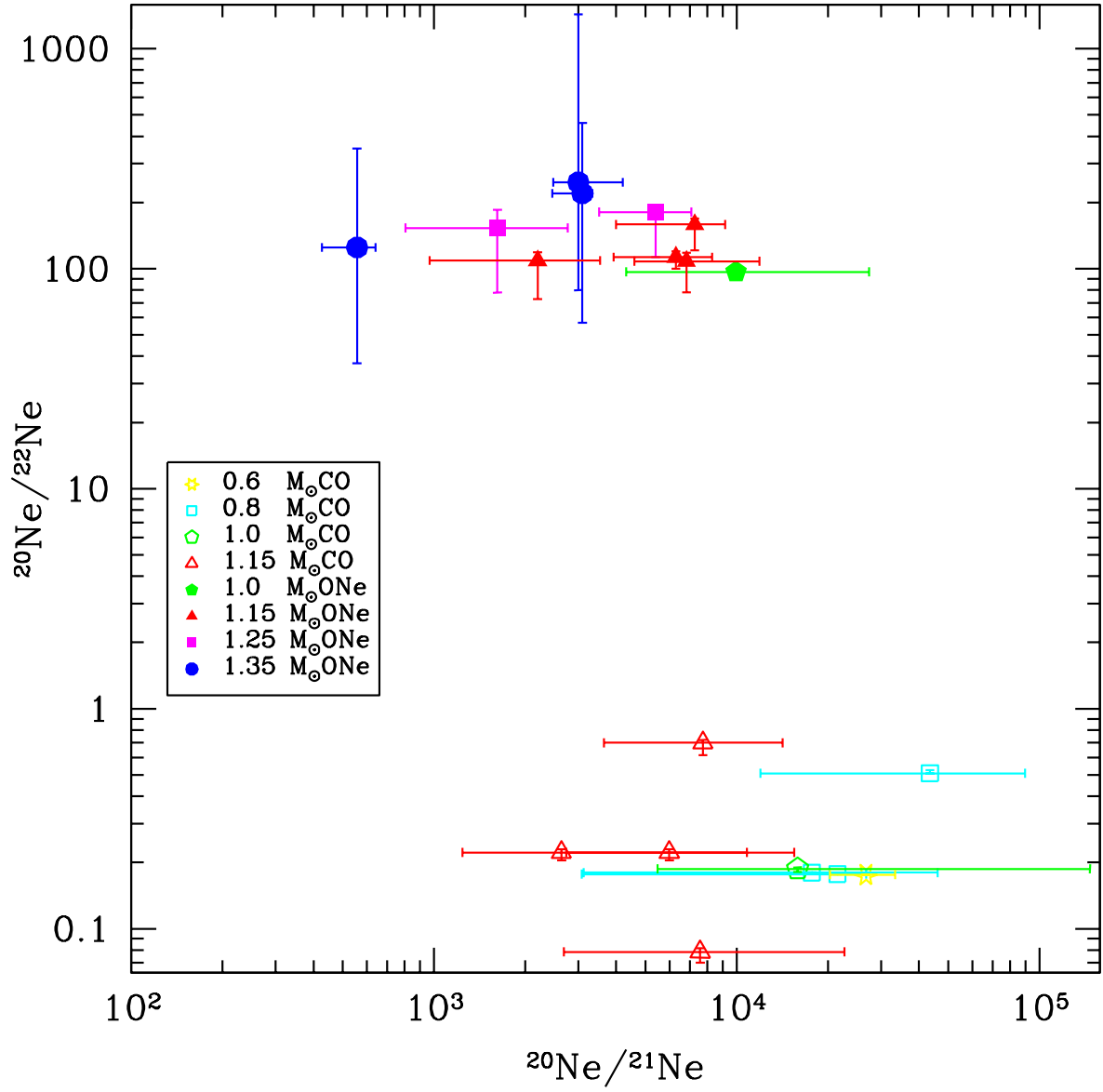}
  \caption{ Same as Fig. 4, for $^{20}$Ne/$^{21}$Ne and $^{20}$Ne/$^{22}$Ne ratios. }
  \end{figure}

  \begin{figure}
  \plotone{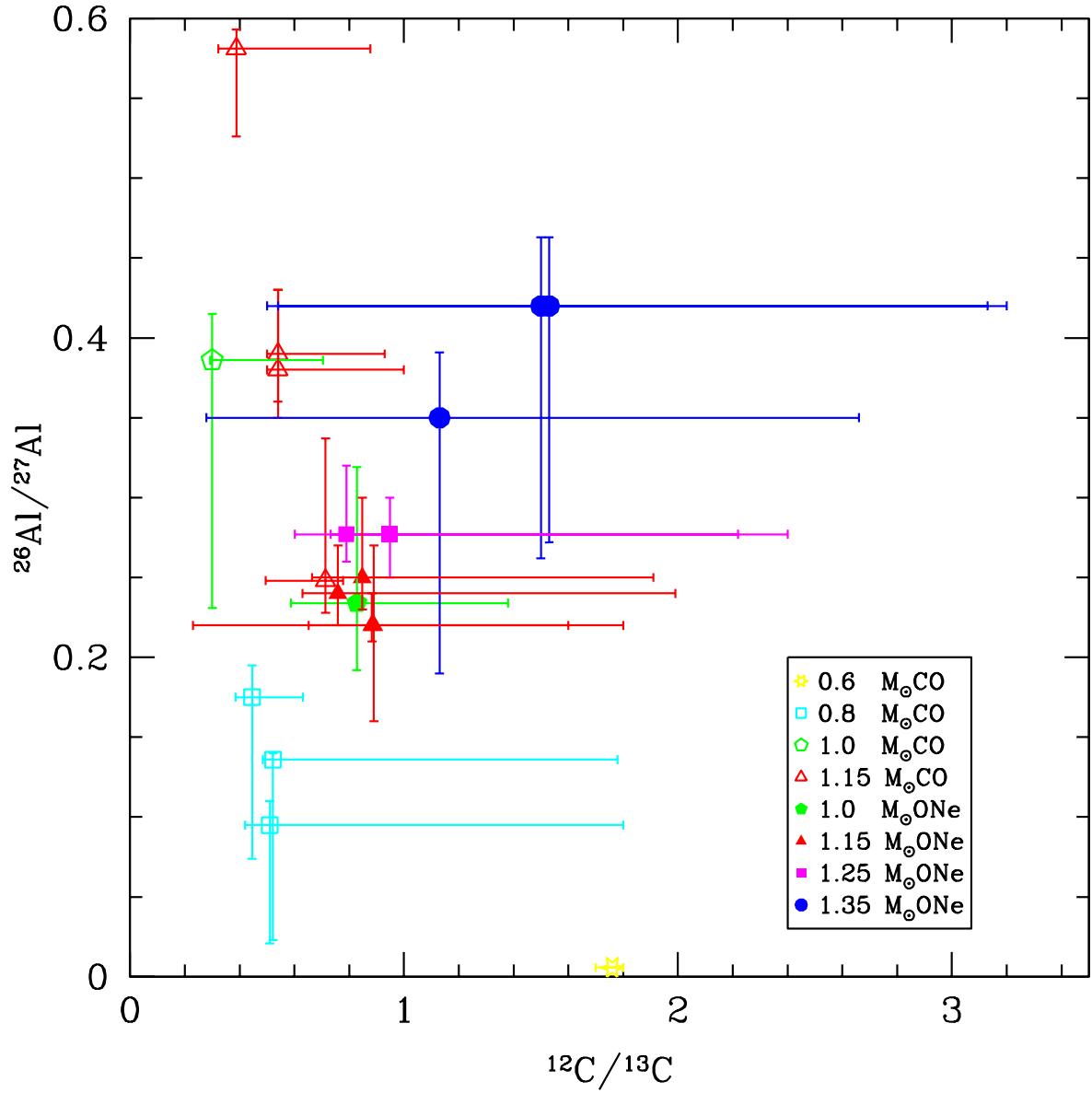}
  \caption{ Same as Fig. 4, for $^{26}$Al/$^{27}$Al versus $^{12}$C/$^{13}$C ratios. }
  \end{figure}

  \begin{figure}
  \plotone{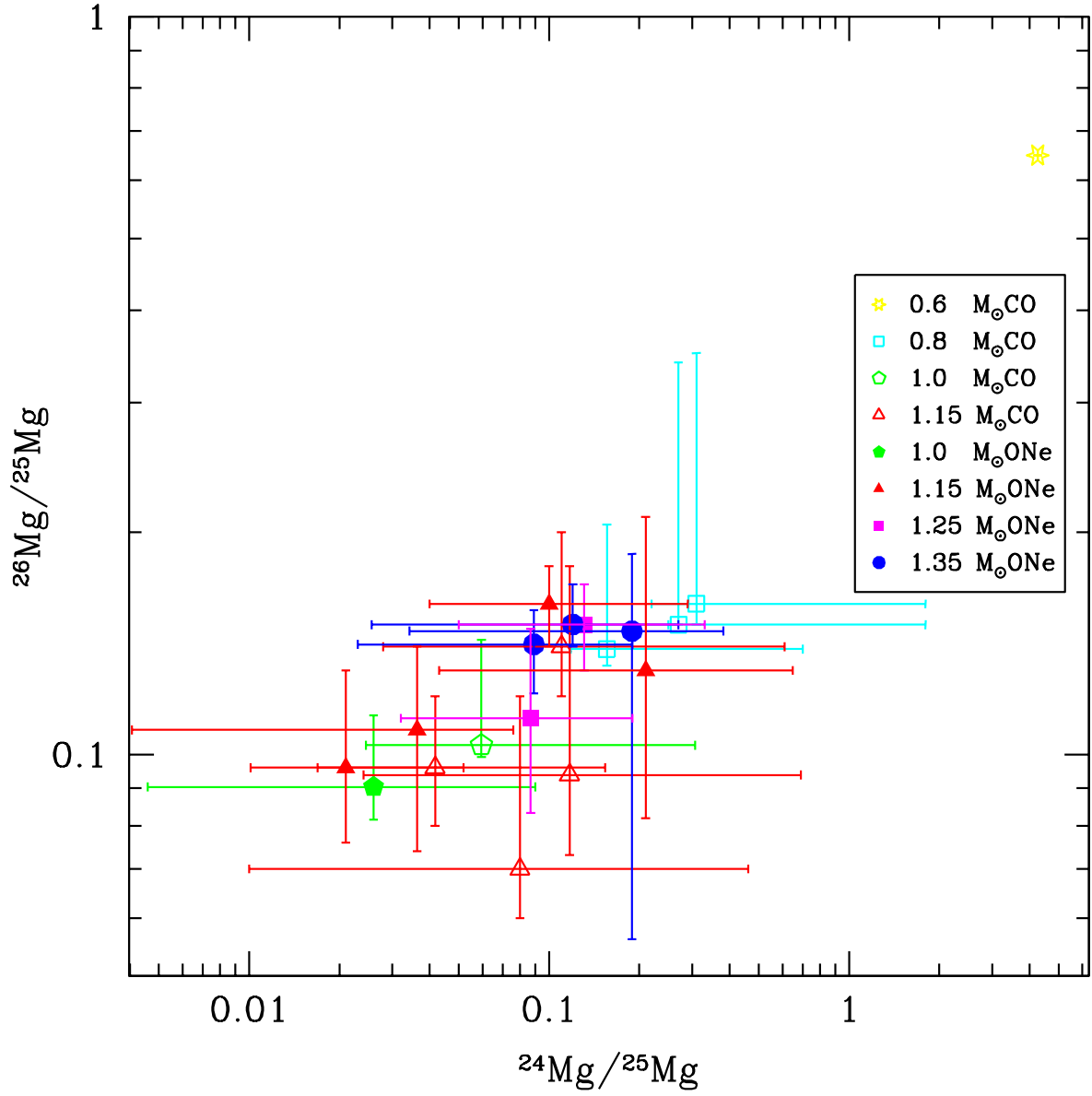}
  \caption{ Same as Fig. 4, for $^{24}$Mg/$^{25}$Mg and $^{26}$Mg/$^{25}$Mg ratios. }
  \end{figure}

\begin{figure}
\plotone{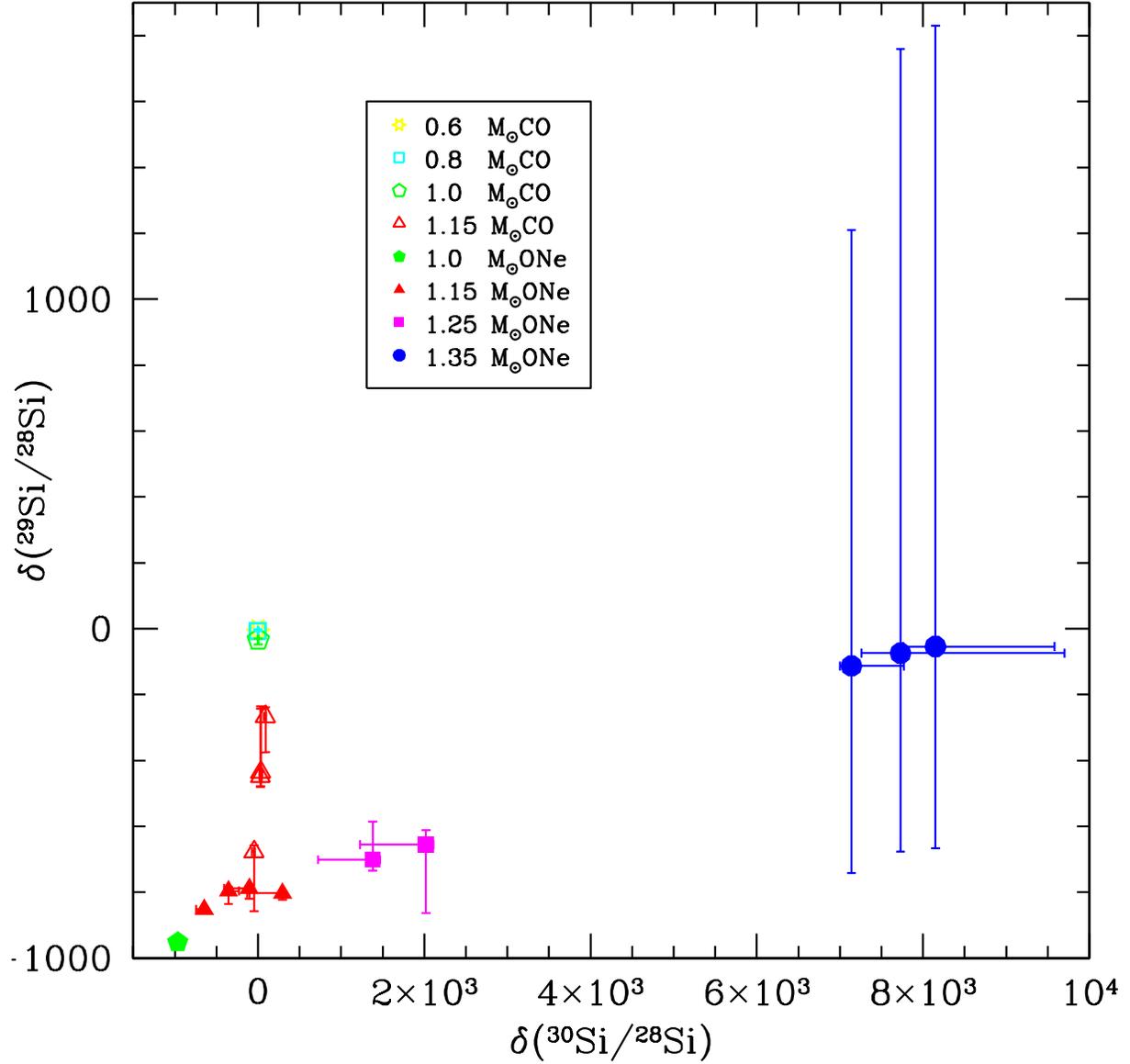}
\caption{Same as Fig. 4, for silicon isotopic ratios,
expressed as delta values (deviations from the solar Si isotopic ratios 
in permil).}
\end{figure}

\begin{figure}
\includegraphics[width=10cm]{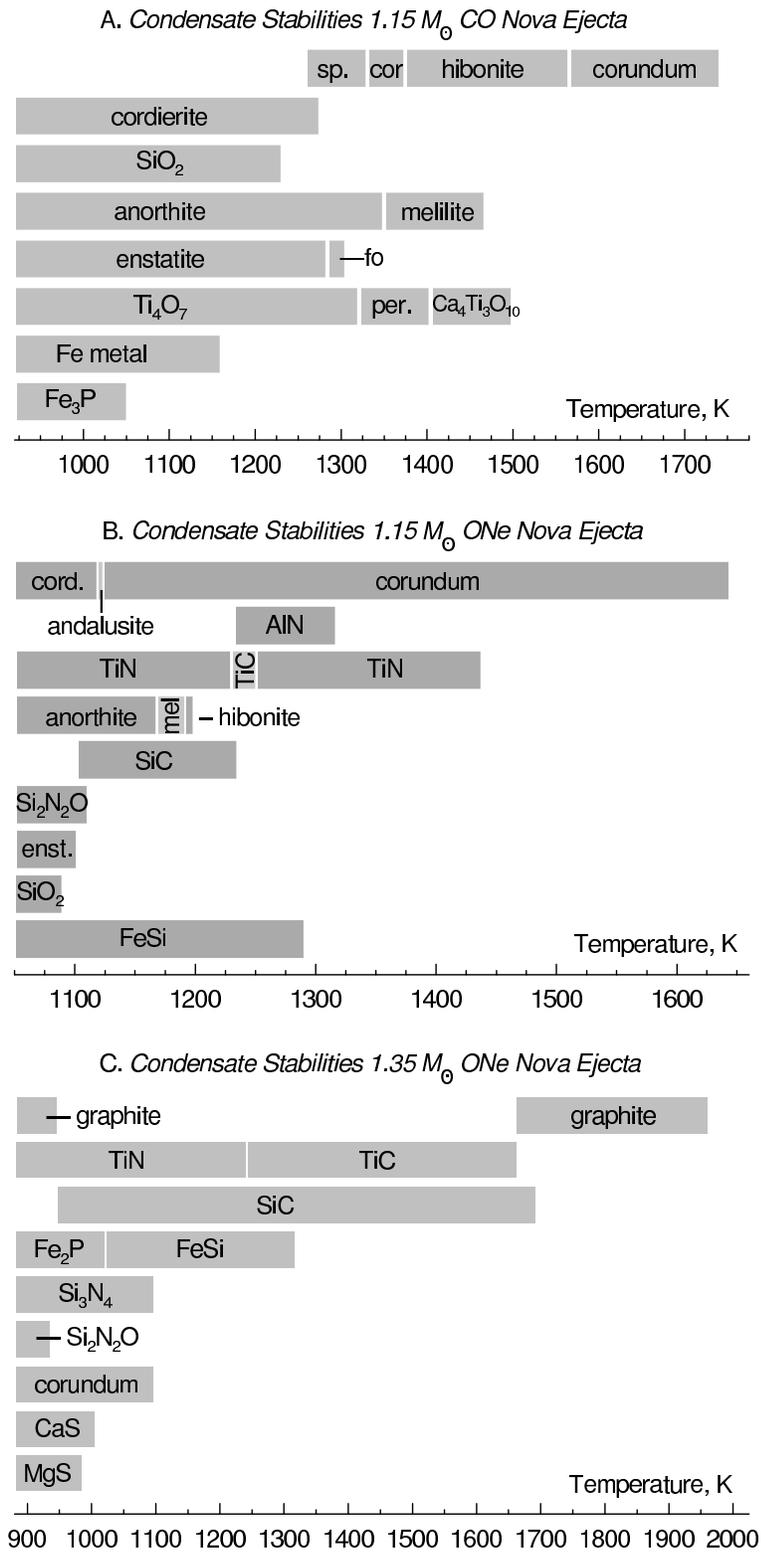}
\caption{Equilibrium condensation sequences showing the different types
of grains expected to form in the ejecta of 3 different classical nova 
outbursts: a 1.15 \msun CO WD (upper panel), a 1.15 \msun ONe WD (middle 
panel), and a 1.35 \msun ONe WD (lower panel).}
\end{figure}

\end{document}